\documentclass[12pt,aps,floatfix]{revtex4}
\usepackage{amsmath}
\usepackage{latexsym}
\usepackage{float}
\usepackage{amssymb}
\usepackage{graphicx}
\usepackage{epsfig}
\begin{document}
\title{From Capillary Condensation to Interface Localization Transitions
in Colloid Polymer Mixtures Confined in Thin Film Geometry}

\author{Andres De Virgiliis$^{1,2}$, Richard L. C. Vink$^3$, J\"urgen Horbach$^4$,
 and Kurt Binder$^1$\\
$^1$ Institut f\"ur Physik, Johannes Gutenberg-Universit\"at Mainz,\\
Staudinger Weg 7, 55099 Mainz, Germany\\
$^2$ Instituto de Investigaciones Fisicoquimicas Teoricas y Aplicadas (INIFTA),
UNPL-CONICET, C.C. 16, Suc.~4, 1900 La Plata, Argentina\\
$^3$ Institut f\"ur Theoretische Physik, Georg August-Universit\"at G\"ottingen,\\
Friedrich-Hund-Platz 1, 37077 G\"ottingen, Germany\\
$^4$Institut f\"ur Materialphysik im Weltraum, Deutsches Zentrum
f\"ur Luft- und Raumfahrt (DLR), 51170 K\"oln, Germany}

\begin{abstract}
Monte Carlo simulations of the Asakura-Oosawa (AO) model for 
colloid-polymer mixtures confined between two parallel repulsive structureless
walls are presented and analyzed in the light of current theories on
capillary condensation and interface localization transitions. Choosing a
polymer to colloid size ratio of $q=0.8$ and studying ultrathin films in
the range of $D=3$ to $D=10$ colloid diameters thickness, grand canonical
Monte Carlo methods are used; phase transitions are analyzed via finite
size scaling, as in previous work on bulk systems and under confinement
between identical types of walls. Unlike the latter work, inequivalent
walls are used here: while the left wall has a hard-core repulsion for
both polymers and colloids, at the right wall an additional square-well
repulsion of variable strength acting only on the colloids is present. We
study how the phase separation into colloid-rich and colloid-poor phases
occurring already in the bulk is modified by such a confinement. When
the asymmetry of the wall-colloid interaction increases, the character
of the transition smoothly changes from capillary condensation-type to
interface localization-type. The critical behavior of these transitions
is discussed, as well as the colloid and polymer density profiles across
the film in the various phases, and the correlation of interfacial
fluctuations in the direction parallel to the confining walls. The
experimental observability of these phenomena also is briefly discussed.
\end{abstract}

\maketitle

\section{INTRODUCTION AND OVERVIEW}
When fluid systems are confined in nanoscopic pores or channels,
one expects that the phase behavior can be profoundly modified
\cite{1,2,3,4,5,6,7,8,9,10}. Such effects have found an increasing
attention recently, for instance because of the current interest
to fabricate devices of nanoscopic size and to manipulate chemical
reactions in nanoscopic reaction volumes (``lab on a chip''), etc.
\cite{11,12,13,14,15,16}. In addition, porous materials with pores
of nanoscopic widths are useful as catalysts or for applications
such as mixture separation, pollution control, etc. \cite{6,
17,18,19}.

However, such applications often are based on empirical knowledge,
the theoretical understanding of confined fluids still being
rather limited \cite{6,7,8,9,10}. In order to make progress with
the theoretical description of fluids under confinement by the
methods of statistical thermodynamics, it is desirable to start
with relatively simple model systems, where both the geometry of
confinement is well characterized, and the relevant interactions
among the fluid particles and between the fluid particles and the
confining solid surfaces are sufficiently well understood. Last
but not least, suitable experimental tools should be in principle
available to put the theoretical predictions to a test.

For these purposes it is hence useful to consider colloidal
suspensions \cite{20,21,22,23,24}, exploiting the analogy between
colloidal fluids and fluids formed from small molecules, but
taking advantage of the much larger length scales (in the $\mu$m
range), of the colloidal particles. Such systems allow detailed
experiments in which individual particles can be tracked through
space in real time using confocal microscopy techniques \cite{25}.
Particularly useful systems in the present context are
colloid-polymer mixtures, which can undergo in the bulk a
liquid-vapor like phase separation into a colloid-rich phase (the
``liquid'') and a colloid-poor phase (the ``vapor'') \cite{23,26}.
This phase separation is due to the (entropic) depletion
attraction between the colloids caused by the polymers. A very simple model,
due to Asakura and Oosawa \cite{27} and Vrij \cite{28}\, describes
the resulting phase separation in the bulk \cite{29,30,31,32,33}
in excellent qualitative agreement with the experiment \cite{23}.
While initially it was thought that mean field theory \cite{29}
accounts very accurately for the Monte Carlo (MC) simulation results
\cite{30,31} of this Asakura-Oosawa (AO) model, a more extensive
MC simulation study \cite{32,33} revealed clear evidence for 
Ising-like critical behavior \cite{34} over a broad regime of 
control parameters.

When such a colloid-polymer mixture is confined by hard walls, also a
depletion attraction of the colloids and the walls occurs \cite{35}
and can cause (in semi-infinite geometry \cite{36,37,38,39,40})
the formation of wetting layers \cite{41,42,43,44,45,46}. Due to the
very low interfacial tension between unmixed phases \cite{47,48,49,50},
thermally activated capillary-wave fluctuations \cite{51,52,53,54,55} are
readily observable in experiment \cite{56} and simulation \cite{50}. The
phase behavior of colloid-polymer mixtures in confinement can be also
studied experimentally. Therefore, this issue has been addressed in
recent computer simulation studies, considering the confinement of
colloid-polymer mixtures by two parallel hard walls a distance $D$
apart \cite{9,57,58,59}. These studies have confirmed the fact that
lateral phase separation in a thin film geometry exhibits a critical
behavior belonging to the class of the two-dimensional Ising model
\cite{58}. Also the the scaling relations of Fisher and Nakanishi \cite{60}
have been verified. Unlike the case of confinement of small molecule
fluids in nanopores, the size of the particles in colloidal fluids by
far exceeds the scale of the atomistic corrugation of the pore walls,
and hence the effects of this corrugation on the packing of particles
near the walls \cite{61,62} need not be considered here.

A very useful aspect of colloidal suspensions is that interactions
among such particles can be tuned by suitable surface treatment \cite{20,21,22,63}.
E.g., a short-range repulsion between colloidal particles often is
created by coating them with a polymer brush \cite{63,64}.
Similarly, one could cancel (partially
or completely) the depletion attraction of colloids towards a hard
wall by coating the latter with a polymer brush, choosing the
grafting density and chain length of these flexible polymers
appropriately. In a colloid-polymer mixture, however, for moderate
chain stretching in the polymer brush the polymers in the solution
still can penetrate into the brush, experiencing hence a much
weaker interaction than the colloidal particles. Only for strongly
stretched chains, as occurring in very dense polymer brushes
\cite{65}, a repulsion of the polymer coils in the solution would
result as well, even if the chemical nature of the polymers in the
solution and in the brush is identical (``autophobicity effect''
\cite{66,67}).

\begin{figure}
\centering
\includegraphics*[width=1.0\textwidth]{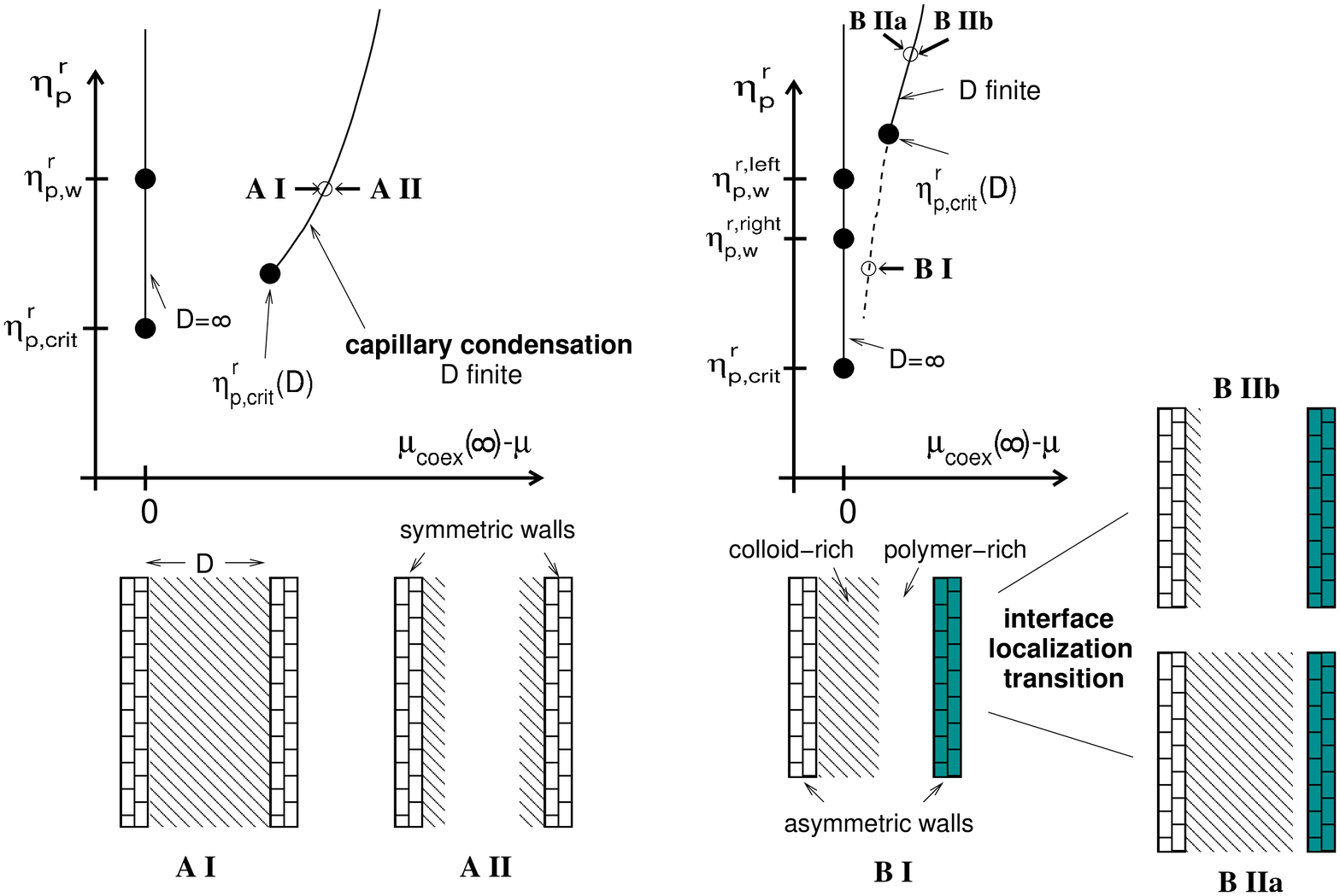}
\caption{\label{fig1} Schematic phase diagrams of a
colloid-polymer mixture confined between two parallel walls a
distance $D$ apart, in the grand-canonical ensemble where the
polymer reservoir packing fraction $\eta_{\rm p}^{\rm r}$ is used as ordinate
axis and the difference between the chemical potential of the
colloids at bulk phase coexistence $\mu_{\textrm{coex}}(D=\infty)$
and the actual chemical potential of the colloids is used as
abscissa (upper part). Phase coexistence in the bulk occurs along
the vertical straight lines at $\mu_{\textrm{coex}}(\infty) - \mu
=0$. The left phase diagram refers to the case of symmetric walls,
the right one to asymmetric walls. The lower part of the figure
indicates the phases that occur in these phase diagrams (shaded 
regions denote colloid-rich domains): In the
case of symmetric walls, a colloid-rich phase (AI) coexists along
the line $\mu=\mu_{\textrm{coex}}(D, \eta_{\rm p}^{\rm r})$ with a
colloid-poor phase (AII). In the case of asymmetric walls, the
analogous phases are BIIa, BIIb; BIIa differs from $A_I$ by the
presence of a layer at the right walls where polymers are
enriched, and BIIb differs from AII by the fact that a surface
enrichment layer of colloids exists only at the left wall.
Finally, a state with a delocalized interface between colloid-rich
and polymer-rich phases (BI) exists along the continuation of the
line $\mu = \mu_{\textrm{coex}}(D,\eta_{\rm p}^{\rm r})$ beyond the critical
point $(\eta_{\rm p, crit}^{\rm r}(D))$, dotted line. The transition from BI to
either BIIa or BIIa when one moves along the dotted line is termed
interface localization transition. For further explanations
cf.~text.}
\end{figure}

This tunability of the wall-colloid interactions opens the
possibility to realize a situation of a slit pore with asymmetric
walls: suppose the left wall is simply a hard wall, attractive for
the colloids, and the right wall a coated hard wall, repulsive for
the colloids (Fig.~\ref{fig1}) \cite{68}. With a colloid-polymer
mixture confined between such asymmetric walls, the possibility
arises to realize the ``interface localization transition''
\cite{7,9,69,70,71,72,73,74,75,76,77,78}. This transition is
illustrated in Fig.~\ref{fig1}. Here, the so-called ``polymer reservoir
packing fraction'' is defined by 
$\eta^{\rm r}_{\rm p} \equiv (4\pi/3)R_{\rm p}^3\exp(\mu_{\rm p}/k_BT)$ 
(with $R_{\rm p}$ and $\mu_{\rm p}$ the radius and the
chemical potential of the polymers, respectively) and plays the role of inverse
temperature when we compare the behavior to that of a fluid of small
molecules that undergoes a liquid-vapor transition. While in the bulk
colloid-polymer mixture phase separation sets in when the variable
$\eta_{\rm p}^{\rm r}$ exceeds the critical value $\eta_{\rm p,
crit}^{\rm r}$, this transition is rounded in the thin film. Starting
out from a layer enriched with colloids on the left wall and enriched
with polymers at the right wall, a stratified domain structure forms,
with a domain wall separating the colloid-rich phase in the left
part and the polymer-rich phase in the right part of the slit pore
(state BI in Fig.~\ref{fig1}). Only at a much larger value $\eta_{\rm
p, crit}^{\rm r}(D)$ a sharp phase transition occurs in the thin film,
with the colloid-polymer interface being bound either to the right wall
(phase B IIa) or to the left wall (phase B IIb). Along the line $\mu =
\mu_{\rm coex}(D, \eta_{\rm p}^{\rm r})$ these two phases may coexist.

Of course, in an experiment one does not have at one's disposal the
intensive variables $\mu$ and the ``polymer reservoir packing fraction''
$\eta_{\rm p}^{\rm r}$, but rather the volume fractions of colloids and 
polymers,
\begin{equation}\label{eq1}
\eta_{\rm c}=\frac{4\pi}{3} 
  R^3_{\rm c} N_{\rm c}/V\;, \quad 
  \eta_{\rm p} = \frac{4\pi}{3} R_{\rm p}^3 N_{\rm p}/V\; ,
\end{equation}
where $V$ is the volume of the system, $R_{\rm c}$ the
radius of the spherical colloidal particles, and $N_{\rm c}$,
$N_{\rm p}$ are the particle numbers of colloids and polymers,
respectively. Since $\eta_{\rm c}$, $\eta_{\rm p}$ are densities of
extensive thermodynamic variables, the first order transition lines
$\mu_{\textrm{coex}}(D,\eta_{\rm p}^{\rm r})$ in the plane of variables
$\eta_{\rm c}$, $\eta_{\rm p}$ are split into two phase coexistence
regions. Bringing the thin film from the one-phase region to inside the
two-phase region (e.g.~by adding polymers to the solution), one creates
a state of the slit pore where in parts of the system the interface is
bound to the left wall and in other parts it is bound to the right
wall. These phases are then separated by interfaces running across
the film from the left to the right wall (or vice versa). A similar
phase coexistence between the two phases AI, AII occurs in the
case of capillary condensation-like transitions for symmetric walls
(left part of Fig.~\ref{fig1}). As always, the amounts of
the coexisting phases is controlled by the lever rule.

In the limit $D \rightarrow \infty$ of the film thickness, we recover
a semi-infinite system and then wetting transitions are expected to
occur, so that, in the symmetric wall case, in the region $\eta^{\rm
r}_{\rm p, crit} < \eta_{\rm p}^{\rm r} < \eta_{\rm p, w}^{\rm r}$ for
$\mu = \mu_{\textrm{coex}}(\infty)$ both walls are (completely) wet,
while for $\eta_{\rm p}^{\rm r} > \eta_{\rm p,w}^{\rm r}$ the walls
are nonwet (``incomplete wetting'' \cite{36,37,38,39,40}). In fact,
the colloid-rich surface enrichment layers indicated for the phase AII
are the precursors of wetting layers that appear when $D \rightarrow
\infty$. Of course, no (infinitely thick \cite{36,37,38,39,40}) true
wetting layer fits into a thin film of finite thickness $D$, and thus
the wetting transition at $\eta_{\rm p}^{\rm r}=\eta_{\rm p,w}^{\rm r}$
(which we have assumed to be of second order \cite{36,37,38,39,40})
is rounded off in the thin film.

For asymmetric walls in the limit $D \rightarrow \infty$ the wetting
transitions at both walls will occur, in general, for different
values of $\eta_{\rm p}^{\rm r}$ at both walls. In Fig.~\ref{fig1}
we have arbitrarily assumed that $\eta_{\rm p, w}^{\rm r, left}
> \eta_{\rm p,w}^{\rm r, right}$. In the simplistic Ising model with
``competing surface magnetic fields'' \cite{69,70,71,72,73,74} $H_1$ and $H_D$,
one can consider a situation with $H_D=-H_1$, where these transitions
then coincide, $\eta_{\rm p,w}^{\rm r, left} = \eta_{\rm p,w}^{\rm r,
right}$. However, such a special symmetry never is expected for a
colloid-polymer mixture (which has an asymmetric phase diagram already
in the bulk). Note, however, that for $D \rightarrow \infty$ one does
not expect that for interface localization transitions $\eta_{\rm p,
crit}^{\rm r}(D)$ converges to the bulk critical point, $\eta_{\rm p,
crit}^{\rm r}$: rather one expects a convergence towards the wetting
transition which is closest to the bulk transition \cite{7}.

In the present paper, we shall present evidence from Monte Carlo
simulations that the scenario sketched in Fig.~\ref{fig1}
is correct, and we shall characterize the behavior of colloid-polymer
mixtures confined by asymmetric walls in detail, considerably extending
preliminary work \cite{68}. Extensive results for the case of
symmetric walls have been presented earlier \cite{58,59}. As in previous
studies in the bulk \cite{32,33} the simulations are carried out mostly
in the grand-canonical ensemble, using a dedicated grand-canonical
cluster algorithm \cite{32} together with re-weighting schemes such as
successive umbrella sampling \cite{79}. Phase transitions are analyzed
by finite size scaling methods \cite{80,81,82}, varying suitably the
lateral linear dimensions $L$ along the walls. For a description of these
techniques, the reader should consult our earlier work \cite{58,59}.

In Sec. II we now present a study of the ``soft mode'' phase \cite{72}
BI for a relatively thick film (thickness $D=10$ colloid diameters). Such
phases with delocalized interfaces are of great interest due to their
large interfacial fluctuations \cite{72,73,74,83,84}, and consequences of
such fluctuations have been seen in experiments both on polymer blends
\cite{85} and colloid-polymer mixtures \cite{46}. Sec. III then gives
a
 discussion of the interface localization transition for an ultrathin
film ($D=3$), attempting to verify the above statement that the critical
exponents should be those of the two-dimensional Ising model. Sec. IV
discusses the phase behavior when both film thickness and the strength
of the short range colloid-wall repulsion are varied. Finally, Sec. V
summarizes some conclusions.

\section{FORMATION AND PROPERTIES OF THE INTERFACE IN THE SOFT
MODE PHASE}

All our Monte Carlo simulations refer to the standard Asakura-Oosawa 
(AO) model and use the same size ratio $q=R_{\rm p}/R_{\rm c}=0.8$ as
the previous work in the bulk \cite{32,33} and for symmetric walls 
\cite{58,59}. In this case, it is known that the critical point in the 
bulk occurs at \cite{32,33}
\begin{equation}\label{eq2}
\eta_{\rm p, crit}^r=0.766 \pm 0.002, \quad
\eta_{\rm c, crit}=0.1340 \pm 0.0002, \quad 
\eta_{\rm p, crit}= 0.3562 \pm 0.0006,
\end{equation}
and also the coexistence curve between the colloid-rich phase
($\eta_{{\rm c},\ell}$) and the polymer-rich phase $(\eta_{{\rm
c},v})$ is known rather precisely, as well as the interfacial tension
\cite{32,33,50}. We now consider a $L \times L \times D$ geometry, where
all lengths are measured in units of the colloid diameter $2R_{\rm c}$,
and periodic boundary conditions are applied in $x$ and $y$-directions
only. For the thickness $D$, the values $D=3$, 5, 7, and 10 are used, 
while the linear dimension $L$ in parallel direction is chosen in the range from $L=15$
to $L=30$. The left wall, located at $z=0$, is taken purely repulsive for
both colloids and polymers. As for the interaction between the colloidal
particles (which is infinite if two colloids overlap and zero else,
as well as the colloid-polymer interaction which also is infinite if
a colloid particle overlaps a polymer and zero else), we take a hard
wall repulsion,
\begin{eqnarray}
U_{\rm w, c}^\ell (z) = \infty, \; z < R_{\rm c}, 
& & U_{\rm w,c}^\ell(z)=0, \; z>R_{\rm c}, \label{eq3} \\
U_{\rm w, p}^\ell (z) = \infty, \; z < R_{\rm p}, 
& & U_{\rm w,p}^\ell(z)=0, \; z>R_{\rm p}, \label{eq4}
\end{eqnarray}
for both colloids $[U_{\rm w,c}^\ell(z)] $ and polymers $[U_{\rm
w,p}^\ell(z)]$. At the right wall, however, we add a square well
potential of strength $\varepsilon$ and with an additional range $R_{\rm
c}$. Thus, the potential acting on the colloids is
\begin{subequations}
\begin{eqnarray}
U_{\rm w,c}^{\rm r}(z)= 0, & & z < D - 2 R_{\rm c}, \label{eq5a}\\
U_{\rm w,c}^{\rm r}(z)=\varepsilon, & & D - 2R_{\rm c}<z<D-R_{\rm c}, 
\label{eq5b}\\
U_{\rm w,c}^{\rm r}(z)=\infty, & & z>D-R_{\rm c}. \label{eq5c}
\end{eqnarray}
\end{subequations}
This square well potential [Eq.~(\ref{eq5b})] could be realized by a
polymer brush of low grafting density and height $R_{\rm c}$, for
instance, so that the region of $z$ where the colloid penetrates into
the brush leads to a finite energy penalty $\varepsilon$ only (note
that we use the convention that the temperature $k_BT=1$; of course,
one could also consider square well potentials of arbitrary range). For
the polymers, on the other hand, the interaction is taken to be of the
same type as in Eq.~(\ref{eq4}),
\begin{equation}\label{eq6}
U_{\rm w,p}^{\rm r}(z)=0, \; z<D-R_{\rm p}, 
\quad U_{\rm w,p}^{\rm r}(z)= \infty, \; z>D-R_{\rm p}.
\end{equation}
This potential models the interactions of polymers with a hard wall
coated with polymer brushes: Under good solvent or Theta solvent
conditions \cite{86}, polymers can overlap with weakly stretched polymer
brushes with little free energy cost.

\begin{figure}
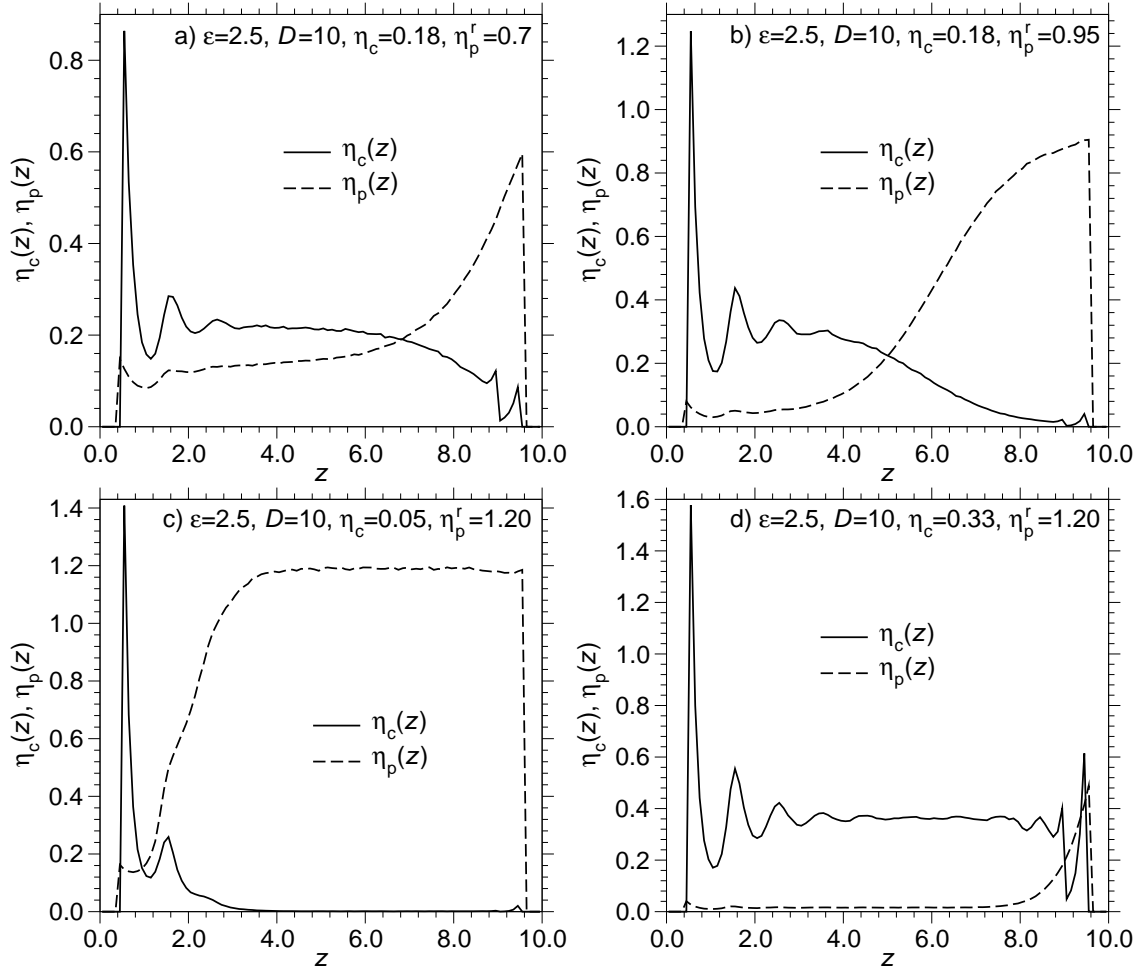

\centering
\includegraphics*[width=0.45\textwidth]{fig2a}
\includegraphics*[width=0.45\textwidth]{fig2b}
\includegraphics*[width=0.45\textwidth]{fig2c}
\includegraphics*[width=0.45\textwidth]{fig2d}
\caption{\label{fig2} Colloid concentration
profiles $\eta_{\rm c}(z)$ and polymer concentration profiles
$\eta_{\rm p}(z)$ as a function of $z$ for a thin film with asymmetric
walls. At $z=0$ there is a hard wall, where the potentials
$U_{\rm w,c}^\ell (z)$ and $U^\ell_{\rm w,p}(z)$ [Eqs.~(\ref{eq2}), (\ref{eq3})]
act on colloids (c) and polymers (p). At $z=D$ there is another
hard wall for both types of particles, with an additional square
well repulsion acting on the colloids only [Eqs.~(\ref{eq3}), (\ref{eq4}),
(\ref{eq5a}), (\ref{eq5b}), (\ref{eq5c})] with a strength $\varepsilon = 2.5$. Profiles were
obtained at $\eta_{\rm c}=0.18$, $\eta_{\rm p}^{\rm r}=0.70$ (a), 
$\eta_{\rm c}=0.18$, $\eta_{\rm p}^{\rm r}=0.95$ (b), 
$\eta_{\rm c}=0.05$, $\eta_{\rm p}^{\rm r}=1.20$ (c), and 
$\eta_{\rm c}=0.33$, $\eta_{\rm p}^{\rm r}=1.20$ (d). For profiles 
(c) and (d), the choices $\eta_{\rm c}=0.05$ and 0.33 roughly 
correspond to the two branches of the coexistence curve in the bulk.}
\end{figure}

It turns out that a phase behavior as sketched in the right part
of Fig.~\ref{fig1} occurs if $\varepsilon \geq 2.5$.
Figure \ref{fig2} presents some typical profiles of the average
local volume fraction of colloids $\eta_{\rm c}(z)$ and polymers
$\eta_{\rm p}(z)$ across the slit pore, for the case $\varepsilon=2.5$ and
$D=10$. Panel (a) shows the profiles for $\eta_{\rm c}=0.18$ and
$\eta_{\rm p}^{\rm r}=0.7$, corresponding to a state point where the bulk
colloid-polymer mixture is still in the one-phase region.
Nevertheless, the profiles of $\eta_{\rm c}(z)$ and $\eta_{\rm p}(z)$ exhibit
pronounced inhomogeneities: the polymer profile $\eta_{\rm c}(z)$ displays
a pronounced peak close to the right wall, and decays with
increasing distance from the right wall to a plateau, almost
independent of $z$, in the regime $3 \leq z \leq 6$. Very close to
the left wall, where the volume fraction of colloids is strongly
enhanced, the concentration of polymers is also inhomogeneous
(indirectly induced by the colloids, since polymers and colloids
must not overlap), before $\eta_{\rm p}(z)$ abruptly decreases to zero
for $z=R_{\rm p}$. The colloid profile $\eta_{\rm c}(z)$ shows a very
pronounced peak close to $z=R_{\rm c}$, on the other hand, which can be
attributed to the depletion attraction of the colloids to the hard
wall. One can recognize a second peak near $z=1.6$ and a weak
third peak near $z=2.5$, these peaks represent the well-known
``layering'' of hard particles near smooth repulsive walls. In the
central part of the thin film, for $3 \leq z \leq 6$, the profile
$\eta_{\rm c}(z)$ is almost flat; thus the surface enrichment of the
colloidal particles at the hard wall is a short range effect. In
the regime near the right walls, where the polymers are attracted,
we recognize first a smooth decrease of $\eta_{\rm c}(z)$ in the range
where the pronounced increase of $\eta_{\rm p}(z)$ sets in. For
$z=D-2R_{\rm c}=9$, where the additional repulsive potential sets in, a
downward step in $\eta_{\rm c}(z)$ occurs, as expected.

It is interesting to contrast the behavior in panel (a), showing
surface enrichment of colloids (left) and polymers (right) at the
walls confining an otherwise homogeneous mixture, with the
behavior in panel (b), which refers to a state where in the bulk
phase separation has occurred. Indeed, Fig.~\ref{fig2}b gives
rather clear evidence for a phase separation in the $z$-direction
perpendicular to the confining walls, of the type denoted as BI in
Fig.~\ref{fig1}. The polymer rich phase occurs on the right side
of the thin film, and $\eta_{\rm p}(z)$ reaches very small values for
$z\leq 4$. Near $z=6$ we recognize inflection points in both
profiles $\eta_{\rm p}(z)$, $\eta_{\rm c}(z)$ as are typical for interfaces
between coexisting phases. Again the profile $\eta_{\rm c}(z)$ exhibits
the typical layering oscillations for small $z$. No such layering
occurs for the polymers near $z=D$, of course, since the
polymer-polymer interaction is zero, the polymer-rich phase is
like a dense ideal gas.

Panels 2c and 2d illustrate states corresponding to the
phases BIIb and BIIa in Fig.~\ref{fig1}, respectively. In the
polymer-rich phase the interface position is at about $z=2.5$, and
unlike Fig.~\ref{fig2}a (where the interface is freely fluctuating
in the center of the slit pore) the width of the interface is only
about two colloid diameters. Such a state is typical for a
colloid-polymer interface tightly bound to the left wall.
Figure \ref{fig2}d is the counterpart showing the profiles in the
colloid-rich phase, where almost all polymers are expelled, apart from
the immediate neighborhood of the right wall.

\begin{figure}
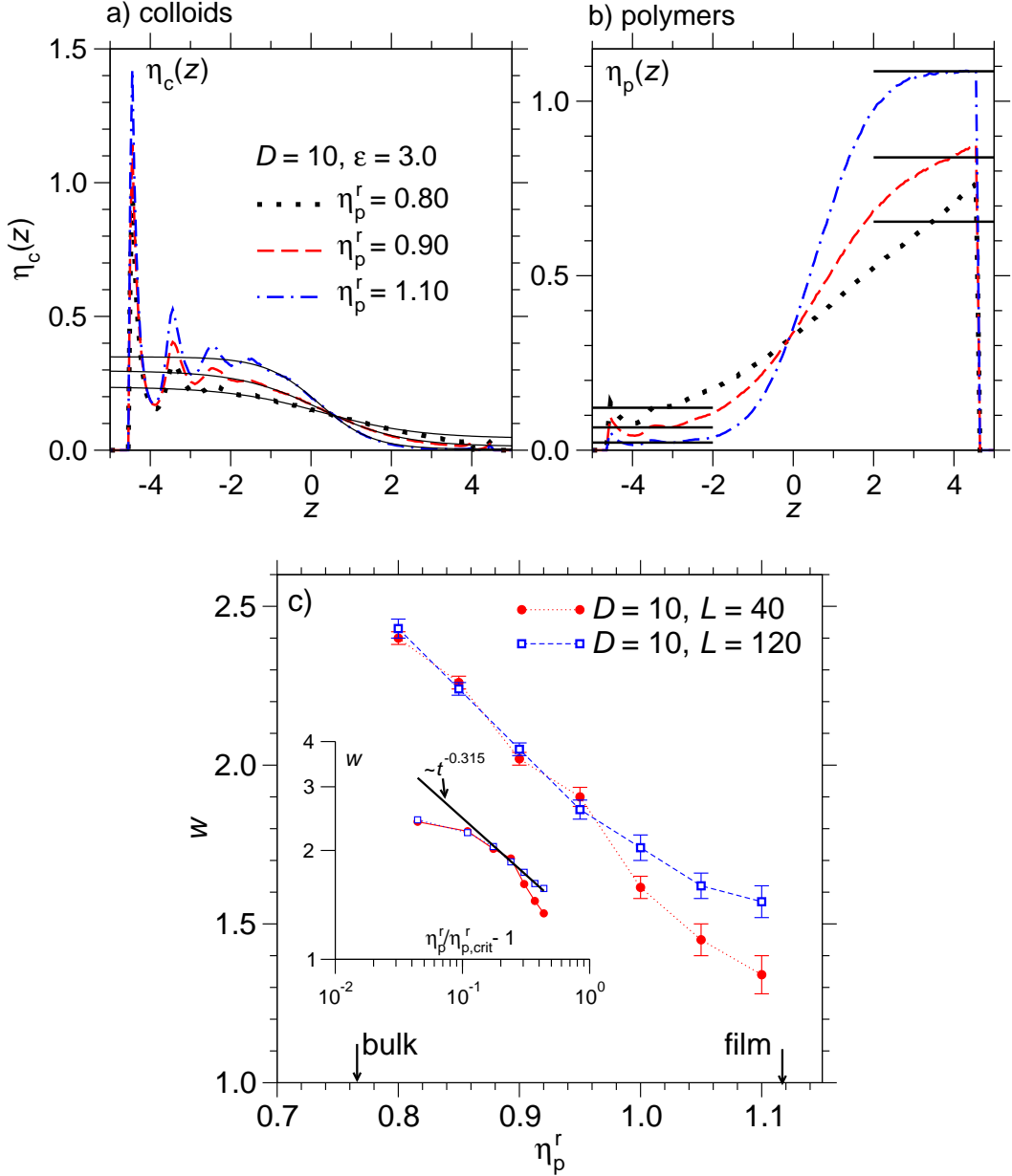

\centering
\includegraphics*[width=0.85\textwidth]{fig3ab}
\vspace*{0.5cm}

\includegraphics*[width=0.55\textwidth]{fig3c}

\caption{\label{fig3} Colloid density profiles $\eta_{\rm c}(z)$
(a) and polymer density profiles $\eta_{\rm p}(z)$ (b)
plotted vs.~$z$ for the case $\varepsilon = 3.0$, $D=10$, and choosing
the average value of $\eta_{\rm c}$ equal to the diameter $\delta$,
$\delta = (\eta_{\rm c}^v + \eta_{\rm c}^\ell)/2$. Here the origin at $z=0$ is
put into the center of the slit pore. Broken curves denote fits
with tanh profiles, as described in the text. Part (c) shows the
dependence of the width $w$ on $\eta_{\rm p}^{\rm r}$. Critical values
$\eta_{\rm p, crit}^{\rm r}$ in the bulk and in the thin film
$\eta_{\rm p, crit}^{\rm r}(D)$ are shown by arrows. Squares are for
the choice $D=10$, $L=120$, circles for $D=10$, $L=40$. The insert
shows a log-log plot of $w$ vs.~$\eta_{\rm p}^{\rm r} 
/\eta_{\rm p, crit}^{\rm r}-1$. The solid line in the insert is a fit with a
power law with the exponent $\nu/2=0.315$ (see text).}
\end{figure}

We conclude that these profiles do give qualitative evidence for
the existence of all three phases BI, BIIa and BIIb in
Fig.~\ref{fig1}. We now study the phase with the delocalized
interface (BI) more closely. In particular, we are interested in how
the interfacial profiles change when the inverse-temperature-like
variable $\eta_{\rm p}^{\rm r}$ is varied (Fig.~\ref{fig3}). Defining an order
parameter $m$ and the coexistence diameter $\delta$ as follows,
\begin{equation}\label{eq7}
m=(\eta_{\rm c}^\ell - \eta_{\rm c}^v)/2, 
\quad \delta = (\eta_{\rm c}^v + \eta_{\rm c}^\ell)/2,
\end{equation}
we choose the average volume fraction of the colloids such that
$\eta_{\rm c}=\delta$, and we attempt to fit the colloid density profile
by a tanh function,
\begin{equation}\label{eq8}
\eta_{\rm c}(z)=\delta - m \tanh[(z-z_0)/w] \quad ;
\end{equation}
here $z_0$ is the position of the interface center and $w$ is the
interfacial width. Fig.~\ref{fig3}a shows that Eq.~(\ref{eq8})
provides a good fit of the colloid density profile, for
all values of $\eta_{\rm p}^{\rm r}$ from 0.90 to 1.10. 
For $\eta_{\rm p}^{\rm r}=0.80$,
however, the profile is extremely wide, due to the proximity of
the critical point in the bulk [Eq.~(\ref{eq1})], and then the fit
is less convincing. Indeed, the polymer density profile
$\eta_{\rm p}(z)$, Fig.~\ref{fig3}b, for $\eta_{\rm p}^{\rm r} = 0.8$ does not even
exhibit an inflection point, while for all larger values of
$\eta_{\rm p}^{\rm r}$ an inflection point clearly is present (it occurs
roughly at $z=z_0$, the inflection point of the polymer density
profile, which is roughly at $z_0 \approx 0.20 \pm 0.05$).

Of course, one notes that
$\eta_{\rm c}(z)$ does not reach the regime of homogeneous ``liquid'' density
$\eta_{\rm c}^\ell$, since for $z \leq -1.5$
in Fig.~\ref{fig3}a the layering effect caused by the repulsive
wall at $z=-5$ already sets in. Likewise, the surface enrichment
of the polymers at the right wall distorts the profiles for $z
\geq 3.5$ in Fig.~\ref{fig3}b. We also note that the profiles seem
to have common intersection points (which do not coincide with
$z_0$, since both $m$ and $\delta$ depend on $\eta_{\rm p}^{\rm r}$). The
common intersection point of the colloid profiles is at $z=0.5 \pm
0.05$, while the common intersection point of the polymer profile
is at $z=0.0 \pm 0.1$. Presumably, these common intersection
points are just numerical coincidences, and will not occur in the
general case (using other choices of $\varepsilon$ and $D$, for
instance). However, the statistical effort for the data in
Fig.~\ref{fig3} is rather substantial, and hence no such
systematic parameter variation has been attempted.

Figure \ref{fig3}c shows that the effective interfacial width $w$
extracted from the fit to Eq.~(\ref{eq8}) increases from about
$w\approx 1.5$ near the critical point of the thin film (the
estimation of thin film critical points is discussed in the
following sections) to about $w\approx 2.4$ for $\eta_{\rm p}^{\rm r}=0.8$.
However, it is important to recall that the width $w$ of the
interface in the ``soft mode'' phase depends on both $\eta_{\rm p}^{\rm r}$
and the total film thickness $D$ \cite{83,84,85,87}. This
complicated behavior results because the ``intrinsic interfacial
profile'' \cite{88,89} is broadened by capillary waves
\cite{51,52,53,54,55}, but the long-wavelength part of the
capillary wave spectrum is suppressed by the effective interface
potential \cite{38,39} caused by the walls. For short range forces
due to the walls, as occurring here, the corresponding prediction for
the mean square width is \cite{83,84,85,87}
\begin{equation}\label{eq9}
w^2=w_0^2[1+\frac{\omega \pi/4}{2+\omega} \frac {D}{w_0}] +\;
\textrm{const}
\end{equation}
Here, $w_0$ is the ``intrinsic width'', which should be related to
the correlation length $\xi_{\rm b}$ along the coexistence curve in the
critical region, $w_0=2 \xi_{\rm b}$, while the wetting parameter
$\omega$ \cite{38,39,40,90,91,92,93} for Ising-like systems is
$\omega \approx 0.8$ and the (unknown) constant due to the short
wavelength cutoff needed in the capillary wave spectrum
\cite{83,84,85} can be neglected near the critical point of the
bulk. The intrinsic width should then vary with 
$\eta_{\rm p}^{\rm r}$ as
\begin{equation}\label{eq10}
w_0=\hat{w}_0(\eta_{\rm p}^{\rm r}/\eta_{\rm p, crit}^{\rm r}-1)^{-\nu},
\quad \quad \nu \approx 0.63,
\end{equation}
with an amplitude factor $\hat{w}_0 $ which is presumably in the range
$0.2 \leq \hat{w}_0 \leq 0.5$ (it is not accurately known since an
unambiguous separation of intrinsic width and capillary wave broadening
is hardly possible in interfacial profiles \cite{50,87}). Since for the
chosen values of $\eta_{\rm p}^{\rm r}$ we have $D \gg w_0$ for $D =
10$ and $(\omega \pi/4)/(2 + \omega) \approx 0.224$, we expect that
$w \approx 1.497 \sqrt{w_0}$ in our case, i.e.~$w$ in Fig.~\ref{fig3}c
should increase with an exponent $\nu/2$. Disregarding the results
for $\eta_{\rm p}^{\rm r}=0.8$ and $\eta_{\rm p}^{\rm r}=0.85$, which are
too close to $\eta_{\rm p, crit}^{\rm r}$ and hence unreliable due 
to finite size effects, we find
that the remaining data for $L=120$ can be nicely fitted to a critical power law
with the expected exponent $\nu/2=0.315$ (see insert of Fig.~\ref{fig3}c).
%but the effective exponent is about $0.4$ rather than $\nu/2$.

\begin{figure}
\centering
\includegraphics*[width=0.95\textwidth]{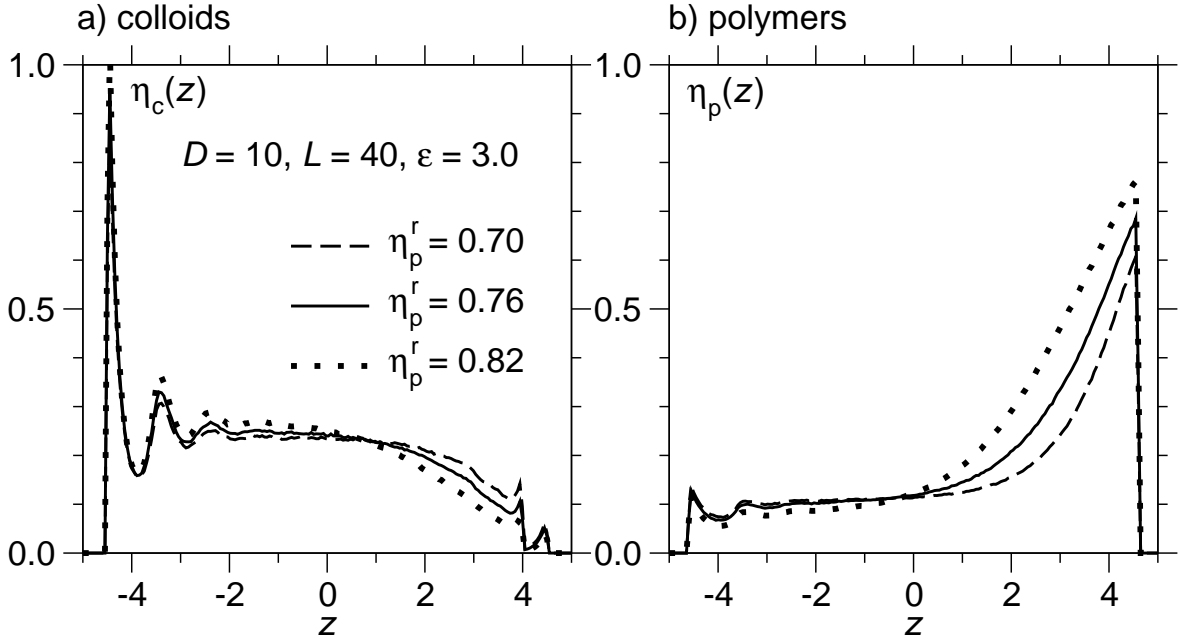}
\caption{\label{fig4} Density profiles $\eta_{\rm c}(z)$
(left part) and $\eta_{\rm p}(z)$ (right part) for the case $D=10$,
$L=40$, $\varepsilon =3$ and $\eta_{\rm c} =0.195$. Three choices of
$\eta_{\rm p}^{\rm r}$ are included, as indicated.}
\end{figure}

Thus, it clearly would be of interest to obtain reliable data
close to the bulk critical point, but then much larger systems
would be required, and this would require very substantial
computer resources, that are not available to us. But we emphasize the fact
that no singular behavior can be observed when at fixed $D$ we
vary $\eta_{\rm p}^{\rm r}$ throughout the bulk critical region, passing the
critical point. As an example, Fig.~\ref{fig4} shows density
profiles for the case $D=10$, $\eta_{\rm c}=0.195, L=40$ and three
values of $\eta_{\rm p}^{\rm r}$ close to $\eta_{\rm p, crit}^{\rm r}$
[Eq.~(\ref{eq1})]. One sees that profiles for $\eta_{\rm p}^{\rm r}$ slightly
above $\eta_{\rm p, crit}^{\rm r}$ and slightly below it are hardly
distinct from each other, all changes with respect to 
$\eta_{\rm p}^{\rm r}$ are very gradual.

\begin{figure}
\centering
\includegraphics*[width=0.95\textwidth]{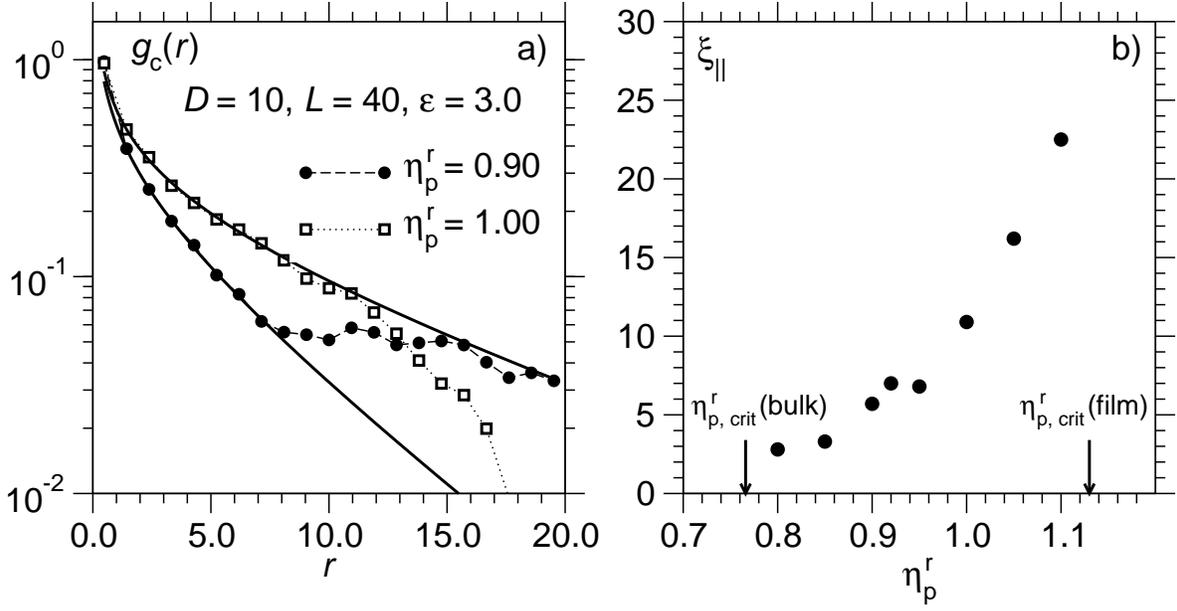}
\caption{\label{fig5} (a) Radial distribution
function $g_{\rm c}(r)$ of the colloidal particles, considering only
distances $r$ parallel to the walls, and particles confined
in the interfacial region $z_0-w<z<z_0+w$, for the case $D=10$,
$L=40$, $\eta_{\rm c}=0.195$, $\varepsilon=3$ and two choices of 
$\eta_{\rm p}^{\rm r}$, as
indicated. Curves are fits to Eq.~(\ref{eq11}). (b) Plot of the
parallel correlation length $\xi_{||}$ extracted from fits as
shown in part (a), versus $\eta_{\rm p}^{\rm r}$.}
\end{figure}

A very interesting property is the correlation function of the
colloidal particles in the interfacial region, $z_0-w<z<z_0+w$ (see
Fig.~\ref{fig5}). If we were to consider an unconfined interface,
the capillary wave fluctuations would cause a power law decay of
these fluctuations. Due to the confinement, the interface feels an
effective potential, and this leads to the existence of a finite
correlation length $\xi_{||}$ of interfacial fluctuations, as
discussed extensively in the literature \cite{72,74,83,84,85,87}.
In simulations of a model for a symmetrical polymer mixture
confined between competing walls, this correlation length was
studied as a function of film thickness. Here we rather study this
quantity as the interface localization transition is approached.
Figure \ref{fig5}a shows that the radial distribution function of
colloidal particles in the interfacial regions is well described
by the formula
\begin{equation}\label{eq11}
g_{\rm c}(r) = \textrm{const} \, \exp (-r/\xi_{||})/\sqrt{r}\quad .
\end{equation}
Equation (\ref{eq11}) was also shown to work very well in the case of
the symmetric polymer mixture \cite{83}. When $\eta_{\rm p}^{\rm r}$
approaches the value $\eta_{\rm p, crit}^{\rm r}(D)$, one sees a strong
increase of $\xi_{||}$, reflecting the expected critical divergence of
$\xi_{||}$ at the interface localization transition [which occurs at
about $\eta_{\rm p, crit}^{\rm r}(D) \approx 1.13 \pm 0.03$]. Arguments
have been given to show that for large enough $D$ there is a region of
mean-field like behavior, where $\xi_{||} \propto (1-\eta_{\rm p}^{\rm
r}/\eta_{\rm p, crit}^{\rm r}(D))^{-\nu_{||}}$ with $\nu_{||}= 1/2$,
while very close to $\eta_{\rm p, crit}^{\rm r}(D)$ the critical behavior
should fall in the class of the two-dimensional Ising model \cite{74},
$\nu_{||}=1$. However, the accuracy of the data in Fig.~\ref{fig5}b
does not warrant an analysis of this crossover behavior.

\section{INTERFACE LOCALIZATION TRANSITION IN VERY THIN FILMS}
\begin{figure}
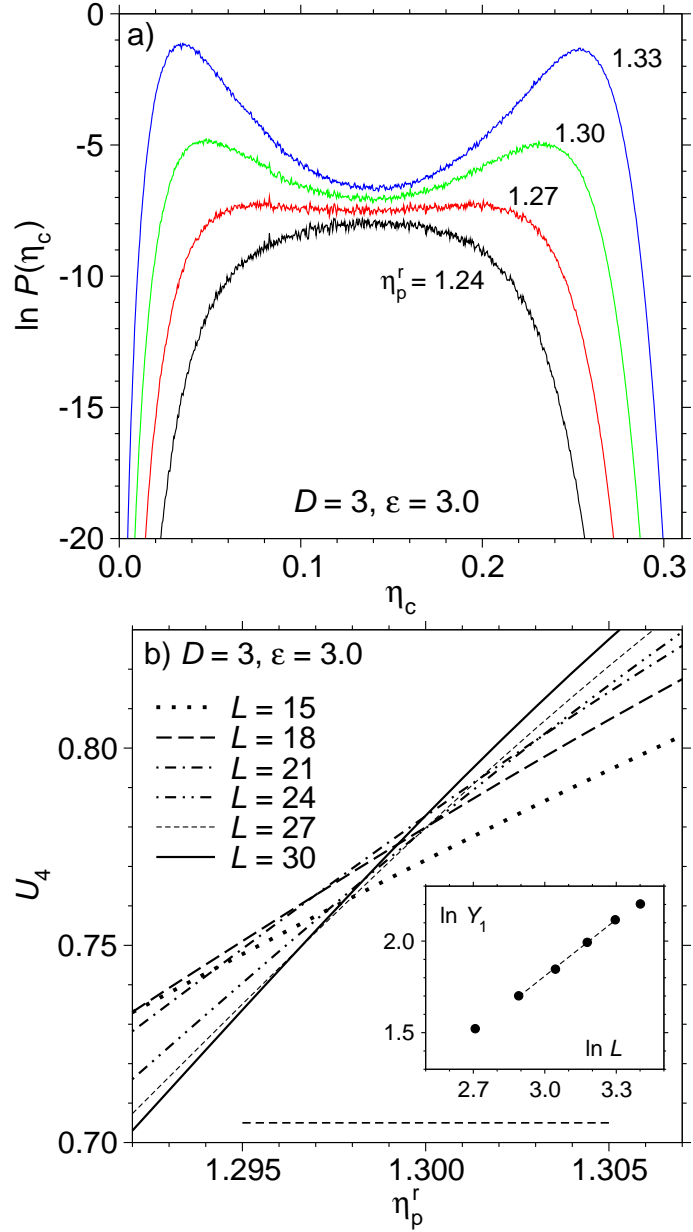

\centering
\includegraphics*[width=0.55\textwidth]{fig6a}
\includegraphics*[width=0.55\textwidth]{fig6b}
\caption{\label{fig6} (a) Logarithm $\ln P(\eta_{\rm c})$
plotted vs.~$\eta_{\rm c}$ for the case $D=3$, $L=15$, $\varepsilon = 3.0$,
and four values of $\eta_{\rm p}^{\rm r}$. (b) $U_4$ plotted vs.~$\eta_{\rm p}^{\rm r}$,
for the case $D=3$, $\varepsilon = 3.0$, and various choices of $L$,
as indicated. The horizontal broken line indicates the range where
intersections occur, $1.295 \leq \eta_{\rm p, crit}^{\rm r}(D) \leq
1.305$. The insert shows a log-log plot of $Y_1=dU_4/d \eta_{\rm p}^{\rm r}$
versus $L$. Broken straight line in the insert illustrates a slope
$1/\nu_{||}=1.035$.}
\end{figure}

Following the procedures used in our earlier study of capillary
condensation in the AO model, we carried out a finite size scaling
analysis of the model with $\varepsilon = 3.0$ for a slit pore which
is only $D=3$ colloid diameters thick. Varying the chemical
potential and applying successive umbrella sampling \cite{79}, the
probability distribution $P(\eta_{\rm c})$ is recorded. Applying
suitable re-weighting techniques \cite{94}, one can apply the
equal area rule \cite{95,96} to determine the chemical potential
$\mu_{\textrm{coex}}$ where the peak of $P(\eta_{\rm c})$ representing
the vapor-like phase and the peak representing the liquid-like
phase have equal weight. Figure \ref{fig6}a shows typical data near
the second order interface localization transition of the thin
film, and Fig.~\ref{fig6}b shows the fourth order cumulant $U_4$
as a function of $\eta_{\rm p}^{\rm r}$ for various $L$ from $L=15$ to $L=30$.
Introducing an order parameter $M$ as $M=\eta_{\rm c} - \langle
\eta_{\rm c}\rangle$, the moments $\langle M^k\rangle$ are defined as
\begin{equation}\label{eq12}
\langle M^k \rangle = \int \limits _0^1 M^k P(\eta_{\rm c}) d\eta_{\rm c},
\end{equation}
and $U_4$ then is given as the ratio of the square of the second
moment and the fourth moment,
\begin{equation}\label{eq13}
U_4=\langle M^2\rangle ^2/\langle M^4\rangle \; .
\end{equation}
For large enough $L$, when finite size scaling \cite{80,81,82}
holds, a convenient recipe to find the critical point
$\eta^{\rm r}_{\rm p, crit}$ is to record $U_4$ for different
choices of $L$ versus $\eta_{\rm p}^{\rm r}$ tuning $\mu$ such 
that $\mu = \mu_{\textrm{coex}}(\eta_{\rm p}^{\rm r})$ and 
look for a common intersection point \cite{80}. For 
$\eta_{\rm p}^{\rm r}\leq \eta_{\rm p, crit}^{\rm r}$ one
fixes $\mu$ by the criterion that $\langle M^2\rangle$ is maximal
\{for $\eta_{\rm p}^{\rm r} > \eta_{\rm p, crit}^{\rm r}$ this 
criterion is an alternative way to estimate 
$\mu_{\rm coex}(\eta_{\rm p}^{\rm r})$\}.

Figure \ref{fig6}a indicates the gradual change from a double peak
distribution to a single peak distribution, which is a
characteristic behavior for all second order phase transitions.
Note that $\eta_{\rm p, crit}^{\rm r}$ does not correspond to the
value of $\eta_{\rm p}^{\rm r}$ where $P(\eta_{\rm c})$ becomes flat over a broad
range of $\eta_{\rm c}$: rather $\eta_{\rm p, crit}^{\rm r}(D)$ still
corresponds to a double peak distribution \cite{80,81,82}.
Figure \ref{fig6}b yields $\eta_{\rm p, crit}^{\rm r}(D=3) = 1.300
\pm 0.005$, i.e.~a value very far away from
$\eta_{\rm p, crit}^{\rm r}$ in the bulk [cf.~Eq.~(\ref{eq1})].
Although it is somewhat disappointing that one cannot really find
a unique intersection point of the cumulants $U_4(\eta_{\rm p}^{\rm r})$ for
the various choices of $L$, one must recognize that for high
enough resolution of the coordinate axes such a scatter is quite
expected, due to residual corrections to finite size scaling
\cite{80}, and due to the statistical errors of the Monte Carlo
data \cite{97}. More disturbing is the fact that the cumulant
intersections occur in a range of values in between the universal
constants $U^*(2 {\rm dim})$ and $U^*(3 {\rm dim})$ for the two- and
three-dimensional Ising model \cite{98,99}, respectively,
\begin{equation}\label{eq14}
U^*(2 {\rm dim})\approx 0.856,\quad U^*(3 {\rm dim}) = 0.629\; .
\end{equation}
As Fig.~\ref{fig6}b shows, intersections occur in the range $0.73
<U^*<0.80$ (although there is some tendency of the intersection
points to move upward with increasing $L$). On the other hand, the
slope of the cumulants at the intersection point, which is
predicted to scale as \cite{80}
\begin{equation}\label{eq15}
dU_4/d\eta_{\rm p}^{\rm r} \propto L^{1/\nu},
\end{equation}
yields an effective exponent rather close to the prediction
$\nu=1$ for the two-dimensional Ising model.

\begin{figure}
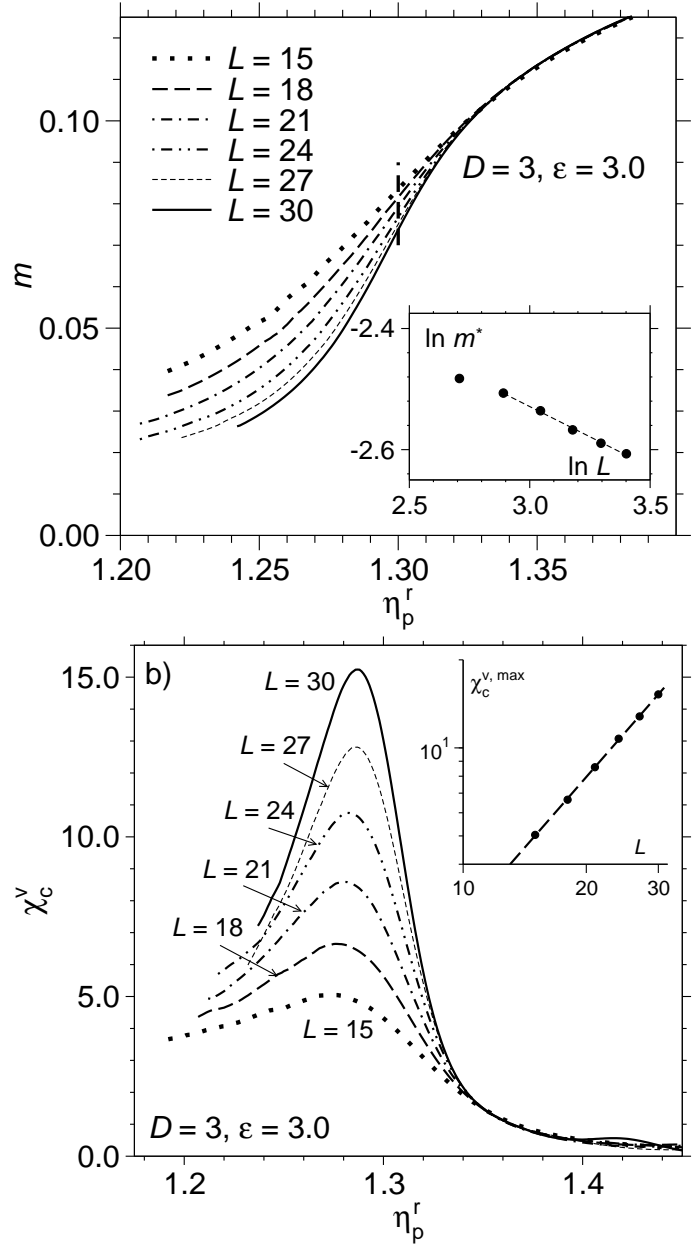

\centering
\includegraphics*[width=0.55\textwidth]{fig7a}
\includegraphics*[width=0.55\textwidth]{fig7b}
\caption{\label{fig7} (a) Order parameter $m$
plotted versus $\eta_{\rm p}^{\rm r}$ for $D=3$, $\varepsilon = 3.0$, and various
choices of $L$ as indicated in the figure. At
$\eta_{\rm p, crit}^{\rm r}$ (indicated by a vertical dashed line)
the resulting data are shown in a log-log plot versus $L$ in the
insert (the broken straight line indicates an effective exponent
$(\beta/\gamma)_{\textrm{eff}} \approx 0.2$). (b) Susceptibility
$\chi_{\rm c}$ plotted versus $\eta_{\rm p}^{\rm r}$ , for the same choices as in
(a). Insert shows the maxima on a log-log plot vs.~$L$ (broken
straight line indicates an exponent $(\gamma/\nu)_{\textrm{eff}}
\approx 1.6)$.}
\end{figure}

\begin{figure}
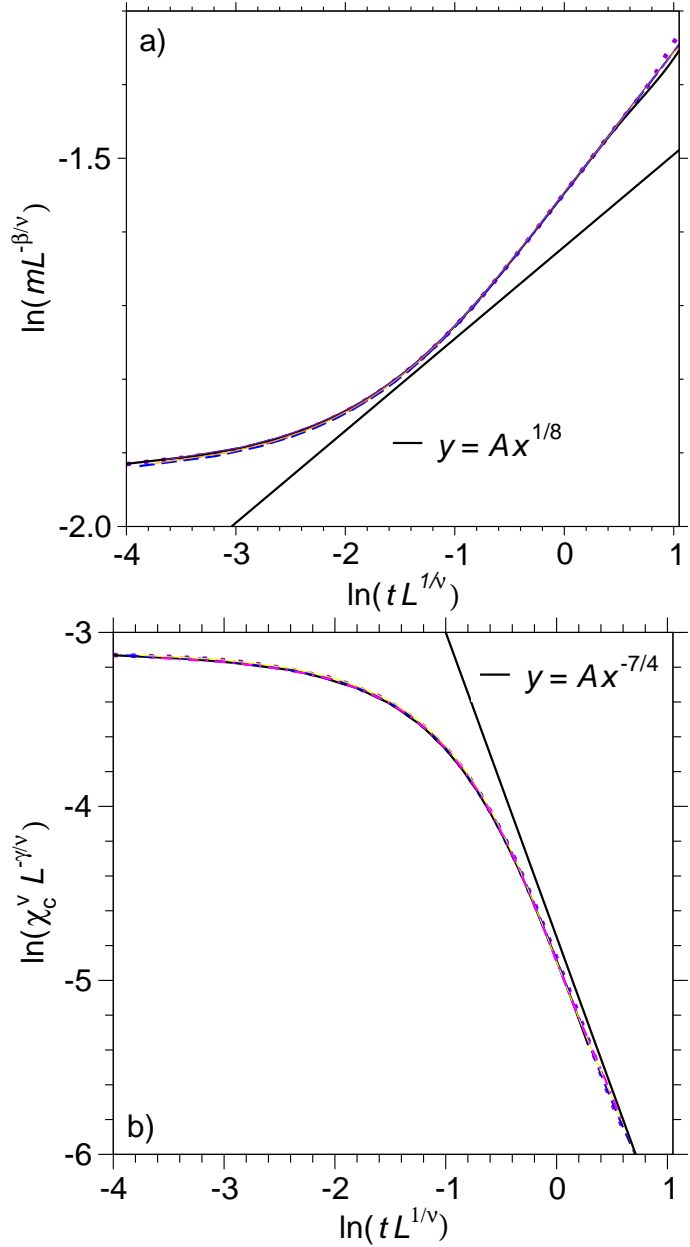

\centering
\includegraphics*[width=0.55\textwidth]{fig8a}
\includegraphics*[width=0.55\textwidth]{fig8b}
\caption{\label{fig8} Scaling plot of the order
parameter (a) and the susceptibility (b), using the three largest
values of $L$ in Fig.~\ref{fig7}, and the effective exponents
(omitting here the index ``eff'') $1/\nu = 1.035,
\beta/\nu=0.2$, and $\gamma/\nu = 1.675$. The straight lines
indicate the theoretical slope of the scaling functions for large
values of $t L^{1/\nu}$, namely $\beta = 1/8$ (a) and $\gamma =
7/4$ (b).}
\end{figure}

Figure \ref{fig7}a shows simulation results for the order parameter
$m=(\eta_{\rm c}^\ell - \eta_{\rm c}^v)/2$, where the volume fractions of
colloids $\eta_{\rm c}^\ell$, $\eta_{\rm c}^v$ are not read off from the peak
positions in Fig.~\ref{fig6}a, since for shallow peaks this would
be a somewhat arbitrary procedure, but rather we take $m$ as the
first moment of the absolute value $\langle |M|\rangle$.
Similarly, Fig.~\ref{fig7}b shows the ``susceptibility''
$\chi_0=L^2D(\langle M^2\rangle - \langle |M|\rangle^2)$. Both
quantities are very strongly affected by finite size effects:
Rather than exhibiting a power law decay, 
$m \propto (1-\eta_{\rm p, crit}^{\rm r}/\eta_{\rm p}^{\rm r})^\beta$ with $\beta = 1/8$, one
finds that approaching $\eta_{\rm p, crit}^{\rm r}$ from above, the
curves for $m$ for the different values of $L$ splay out and
develop very pronounced ``finite size tails'' \cite{80,95} for
$\eta_{\rm p}^{\rm r}<\eta_{\rm p, crit}^{\rm r}$. At
$\eta=\eta_{\rm p, crit}^{\rm r}$ one finds that the data are
compatible with a power law decay (insert to Fig.~\ref{fig7}a)
\begin{equation}\label{eq16}
m^* (L) \equiv m(L,\eta_{\rm p, crit}^{\rm r}) \propto
L^{-\beta/\nu}\; .
\end{equation}
According to the two-dimensional Ising model, one would expect
$\beta/\nu = 1/8$. However, the straight line in the insert of
Fig.~\ref{fig7}a rather indicates an effective exponent of
$(\beta/\nu)_{\textrm{eff}}\approx 0.20 \pm 0.02$. Likewise, the
susceptibility maxima, which should scale as \cite{80,81,82}
\begin{equation}\label{eq17}
\chi_{\rm c}^{\textrm{max}}\propto L^{\gamma /\nu}\;,
\end{equation}
with the two-dimensional Ising value being $\gamma/\nu = 1.75$,
rather suggest an effective exponent $(\gamma/\nu)_{\textrm{eff}}
= 1.60 \pm 0.03$. Very roughly, these exponents are compatible
with the hyperscaling relation \cite{34} $\gamma/\nu + 2
\beta/\nu=2$. Using the quoted effective exponents $1/\nu_{\rm eff}$,
$(\beta/\nu)_{\textrm{eff}}$ and $(\gamma/\nu)_{\textrm{eff}}$,
one finds that on a scaling plot, where the variable $t\equiv
|\eta_{\rm p, crit}^{\rm r}/\eta_{\rm p}^{\rm r}-1|$ is rescaled with
$L^{1/\nu}$ and $m$ or $\chi$ are rescaled with $L^{\beta/\nu}$ or
$L^{-\gamma/\nu}$, one finds reasonable data collapse (Fig.~\ref{fig8}).
Such a partial success of a finite size scaling analysis, i.e.~good
data collapse is only found when effective exponents are used that
deviate somewhat from the theoretical values, has already been
seen for interface localization-delocalization transitions in the
Ising model \cite{73,74} and hence these problems are not a
surprise in the present case.

\section{OVERVIEW OF THE PHASE BEHAVIOR}
\begin{figure}
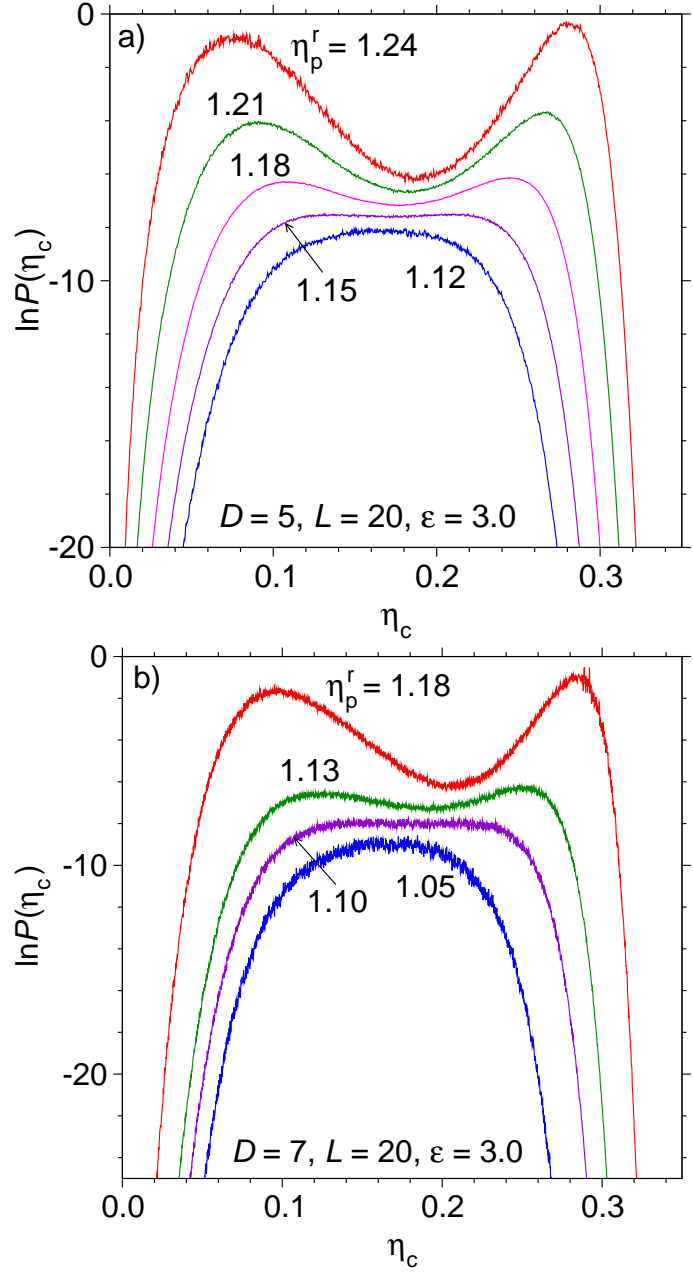

\centering
\includegraphics*[width=0.55\textwidth]{fig9a}
\includegraphics*[width=0.55\textwidth]{fig9b}
\caption{\label{fig9} (a) Logarithm of the
probability distribution of the colloid volume fraction, $\ln
P(\eta_{\rm c})$, plotted vs.~$\eta_{\rm c}$ for $D=5$, $L=20$, $\epsilon=
3.0$, and four choices of $\eta_{\rm p}^{\rm r}$, as indicated. (b) Same as
(a), but for $D=7$.}
\end{figure}

\begin{figure}
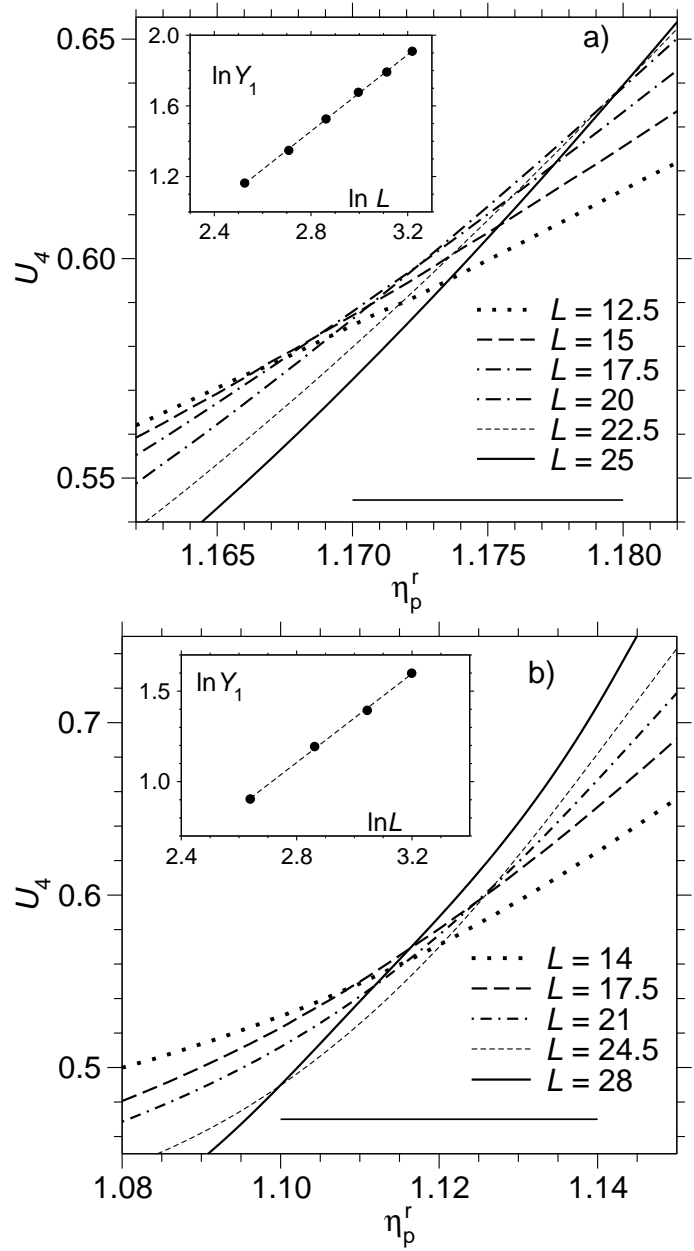

\centering
\includegraphics*[width=0.55\textwidth]{fig10a}
\includegraphics*[width=0.55\textwidth]{fig10b}
\caption{\label{fig10} Cumulant $U_4$ plotted vs.~$\eta_{\rm p}^{\rm r}$ 
for various choices of $L$ as indicated, for $\varepsilon=3.0$ and 
the film thickness $D=5$ (a) and $D=7$ (b). Inserts
show plots of $\ln Y_1$ vs.~$\ln L$, where 
$Y_1=\partial U_4/\partial \eta_{\rm p}^{\rm r}$. The broken 
straight lines indicate the
effective exponent $(1/\nu)_{\textrm{eff}}$, with (a)
$(1/\nu)_{\textrm{eff}} \approx 1.08$ and (b)
$(1/\nu)_{\textrm{eff}} \approx 1.23$. The horizontal straight
lines indicate the accuracy, with which
$\eta_{\rm p, crit}^{\rm r}(D)$ can be estimated, namely 
$\eta_{\rm p, crit}^{\rm r}(D=5)=1.175 \pm 0.005$ and
$\eta_{\rm p, crit}^{\rm r}(D=7) = 1.12 \pm 0.02$.}
\end{figure}

We now describe some of our results for other film thicknesses
$D$. In principle, the same type of analysis was carried out for
$D=5$ and $D=7$ as well, but it turned out that the distribution
$P(\eta_{\rm c})$ for $\eta_{\rm c}> \eta_{\rm c, crit}(D)$ becomes
increasingly asymmetric when $D$ gets larger (Fig.~\ref{fig9}). Also
the cumulant intersections get spread out over a rather large
range of $\eta_{\rm p}^{\rm r}$ (Fig.~\ref{fig10}), and these intersection points lie even in a
range that is below the three-dimensional Ising value,
Eq.~(\ref{eq14}). We interpret this finding as an indication that
with $D$ getting larger an increasing fraction of the critical
region falls into the region of mean-field like behavior, as was
theoretically predicted \cite{74}.

\begin{figure}
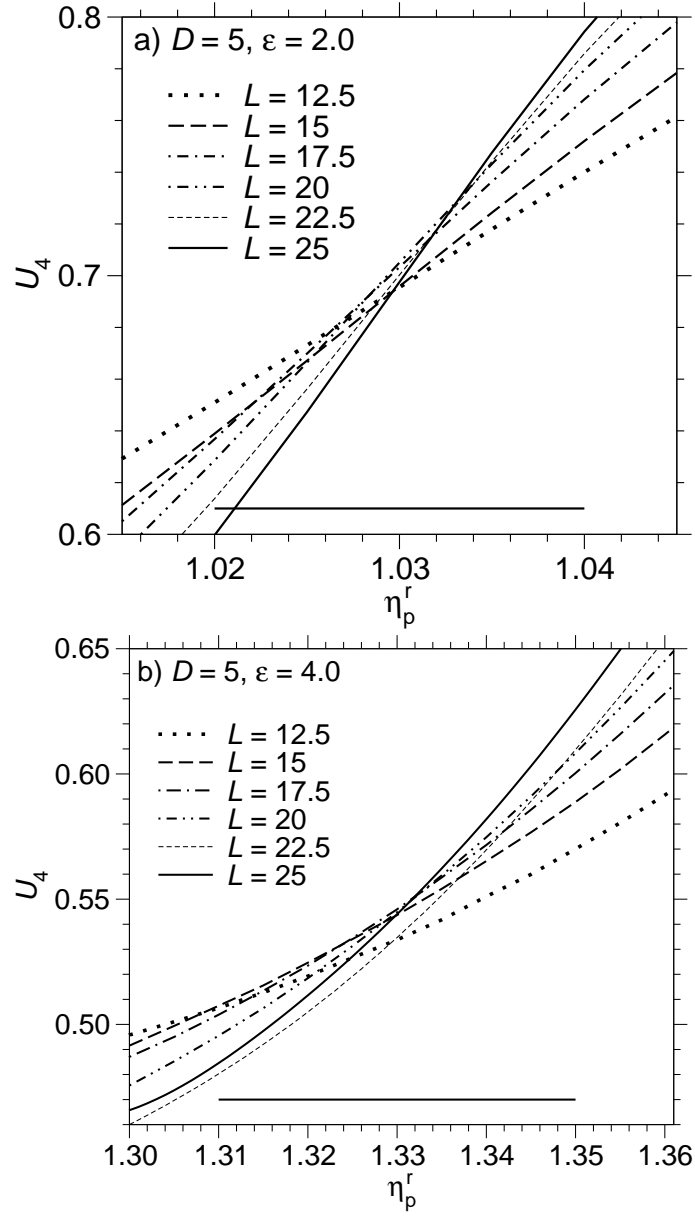

\centering
\includegraphics*[width=0.55\textwidth]{fig11a}
\includegraphics*[width=0.55\textwidth]{fig11b}
\caption{\label{fig11} Cumulant $U_4$ plotted vs.
$\eta_{\rm p}^{\rm r}$ for various choices of $L$ as indicated, for film
thickness $D=5$ and two choices of $\epsilon$, $\varepsilon=2.0$ (a)
and $\varepsilon = 4.0$ (b). The horizontal straight lines
indicate the accuracy with which $\eta_{\rm p, crit}^{\rm r}(D=5)$
can be estimated, namely $\eta_{\rm p, crit}^{\rm r} = 1.03 \pm 0.01$
($\varepsilon = 2.0$) and $\eta_{\rm p, crit}^{\rm r} = 1.33 \pm 0.02$
($\varepsilon = 4$).}
\end{figure}

Also for fixed $D$ the accuracy, with which
$\eta_{\rm p, crit}^{\rm r}(D)$ can be estimated, clearly
deteriorates when $\varepsilon$ increases (Fig.~\ref{fig11}). Note
that data for $D=5$ and $\varepsilon = 1.0$ were already given in our
preliminary communication \cite{68}, the choice $\varepsilon = 1.0$
corresponds to a capillary condensation-type behavior, however.

\begin{figure}
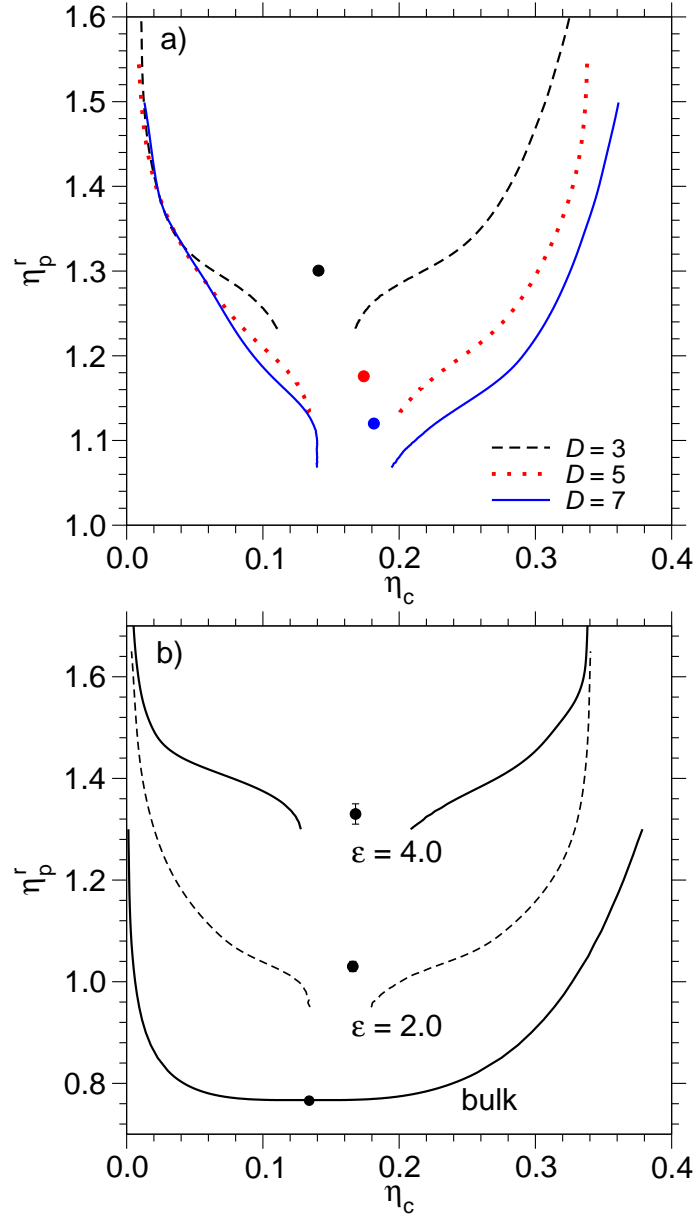

\centering
\includegraphics*[width=0.55\textwidth]{fig12a}
\includegraphics*[width=0.55\textwidth]{fig12b}
\caption{\label{fig12} Phase diagrams in the plane
of variables $\eta_{\rm p}^{\rm r}$ and $\eta_{\rm c}$ for the 
case $L=30$, $\varepsilon=3.0$, and three choices of $D$, $D=3$, 5, and 7 (a),
as well as for the case $D=5$, $L=30$ and several choices of $\varepsilon$ (b).
The full curves show the estimates for the
binodal, while the filled circles represent the estimates of the critical points.
%the broken (almost vertical) curves show the estimates for the coexistence diameter. 
In case (b), the estimate
for the bulk coexistence curve (extrapolated towards $L \rightarrow \infty$) 
is included.}
\end{figure}

\begin{figure}
\centering
\includegraphics*[width=0.95\textwidth]{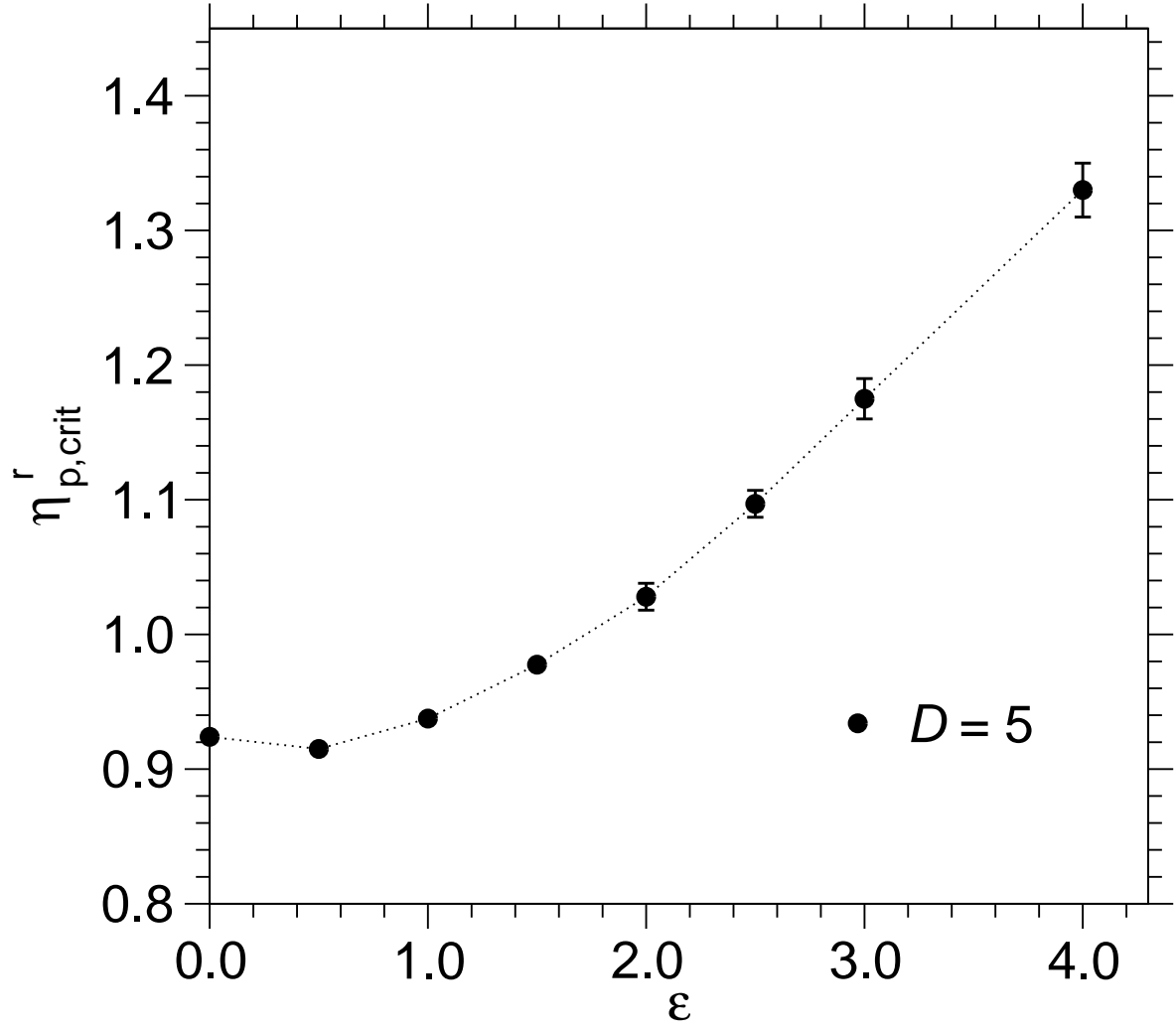}
\caption{\label{fig13} Plot of
$\eta_{\rm p, crit}^{\rm r}(D=5)$ vs.~$\epsilon$.}
\end{figure}

Figure \ref{fig12}a shows estimates for the phase diagrams for
the interface localization transition for $\varepsilon =3$ and three
choices of $D$, while Fig.~\ref{fig12}b shows analogous data for
$D=5$ but varying $\varepsilon$, and Fig.~\ref{fig13} shows a plot
of $\eta_{\rm p, crit}^{\rm r}(D=5) $ vs.~$\varepsilon$. One sees that
miscibility is enhanced if either $D$ decreases, or $\varepsilon$
increases, or both. 
%Note that the coexistence curve data for the
%thin films in Fig.~\ref{fig12} are for one particular choice of
%$L$ only, and are hence near $\eta_{\rm p, crit}^{\rm r}(D)$
%severely affected by finite size effects: the ``finite size
%tails'' cross the horizontal lines that indicate the respective
%values of $\eta_{\rm p, crit}^{\rm r}(D)$. What rather must happen
%in the thermodynamic limit, is that both branches of the
%coexistence curve must bend over before
%$\eta_{\rm p, crit}^{\rm r}(D)$ is reached and must ultimately
%smoothly merge in the point $\eta_{\rm p, crit}^{\rm r}(D)$,
%$\eta_{\rm c, crit}(D)$, where the coexistence diameter
%crosses the line $\eta_{\rm p, crit}^{\rm r}(D)$.

Finally we turn to the variation of $\eta_{\rm p, crit}^{\rm r}$
with $\varepsilon$ for the choice $D=5$ (Fig.~\ref{fig13}). As found from a
self-consistent field calculation for a symmetrical polymer
mixture confined between competing walls \cite{77}, the minimum of
the curve $\eta_{\rm p, crit}^{\rm r}$ does not occur for the case
of symmetric walls ($\varepsilon = 0$), but for an asymmetric
situation. It also is remarkable and unexpected, that for large
$\varepsilon$ the curve for $\eta_{\rm p, crit}^{\rm r}$ does 
not level off.

\begin{figure}
\centering
\includegraphics*[width=0.95\textwidth]{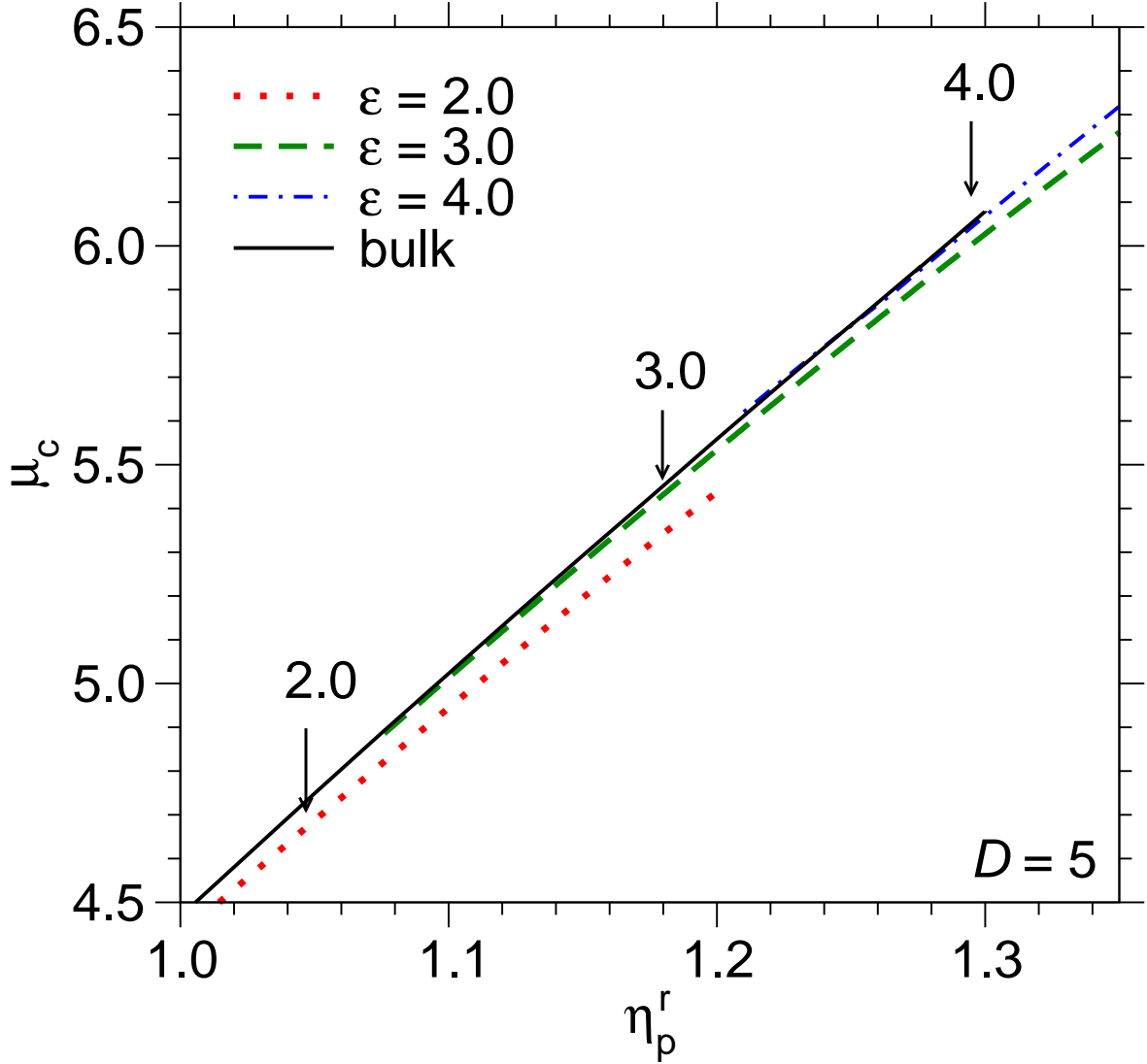}
\caption{\label{fig14} Plot of the colloid chemical potential versus
$\eta_{\rm p}^{\rm r}$. The coexistence curve for asymmetric thin
films, for $D=5$ and three values of $\varepsilon$, $\varepsilon=2.0, 3.0$
and 4.0, as indicated, are compared to the coexistence curve of the
colloid-polymer mixture in the bulk (solid line). Arrows show the
critical points $\eta_{\rm p, crit}^{\rm r}(D)$. Note that the bulk
critical point $(\eta_{\rm p, crit}^{\rm r}=0.766)$ is far beyond the
scale of the diagram. For $\eta_{\rm p}^{\rm r}<\eta_{p,\textrm{crit}}^r(D)$ 
the meaning of the shown
curves denotes the coexistence of domains in the stratified
structure, with a single interface parallel to the walls, located
via the maximum of $\langle M^2 \rangle$.}
\end{figure}

Figure \ref{fig14} shows the counterpart of the schematic Fig.~\ref{fig1}
(left part), presenting in the plane of variables $\mu_{\rm c}$ and
$\eta_{\rm p}^{\rm r}(D)$ the numerical results for the coexistence
curves between colloid-rich and polymer-rich phases, for the case of $2.0
\leq \varepsilon \leq 4.0$, i.e.~the region where interface-localization
transitions occur (which are highlighted in the diagram by arrows). Note
that unlike Fig.~\ref{fig1}, $\mu_{\rm c}(\infty)$ was not subtracted
from $\mu_{\rm c}$, thus the bulk coexistence is not simply the
ordinate axis as in Fig.~\ref{fig1}, but rather a nontrivial curve
(which actually is not very different from a straight line). While
for $\varepsilon = 2.0$ there is still a small but systematic offset
between the curves $\mu_{\rm c}(\eta_{\rm p}^{\rm r}, D=5)$ and
$\mu_{\rm c}(\infty)$, for $\varepsilon = 3.0$ and $\varepsilon =
4.0$ the offset is almost negligibly small. The part of the curves
$\mu_{\textrm{coex}}(\eta_{\rm p}^{\rm r}(D))$ to the left of $\eta_{\rm
p, crit}^{\rm r}(D)$ represents the state BI in the schematic phase
diagram, Fig.~\ref{fig1}, where a delocalized interface occurs in the
center of the film, separating the colloid-rich phase adjacent to the
left wall and the polymer-rich phase adjacent to the right wall.

\begin{figure}
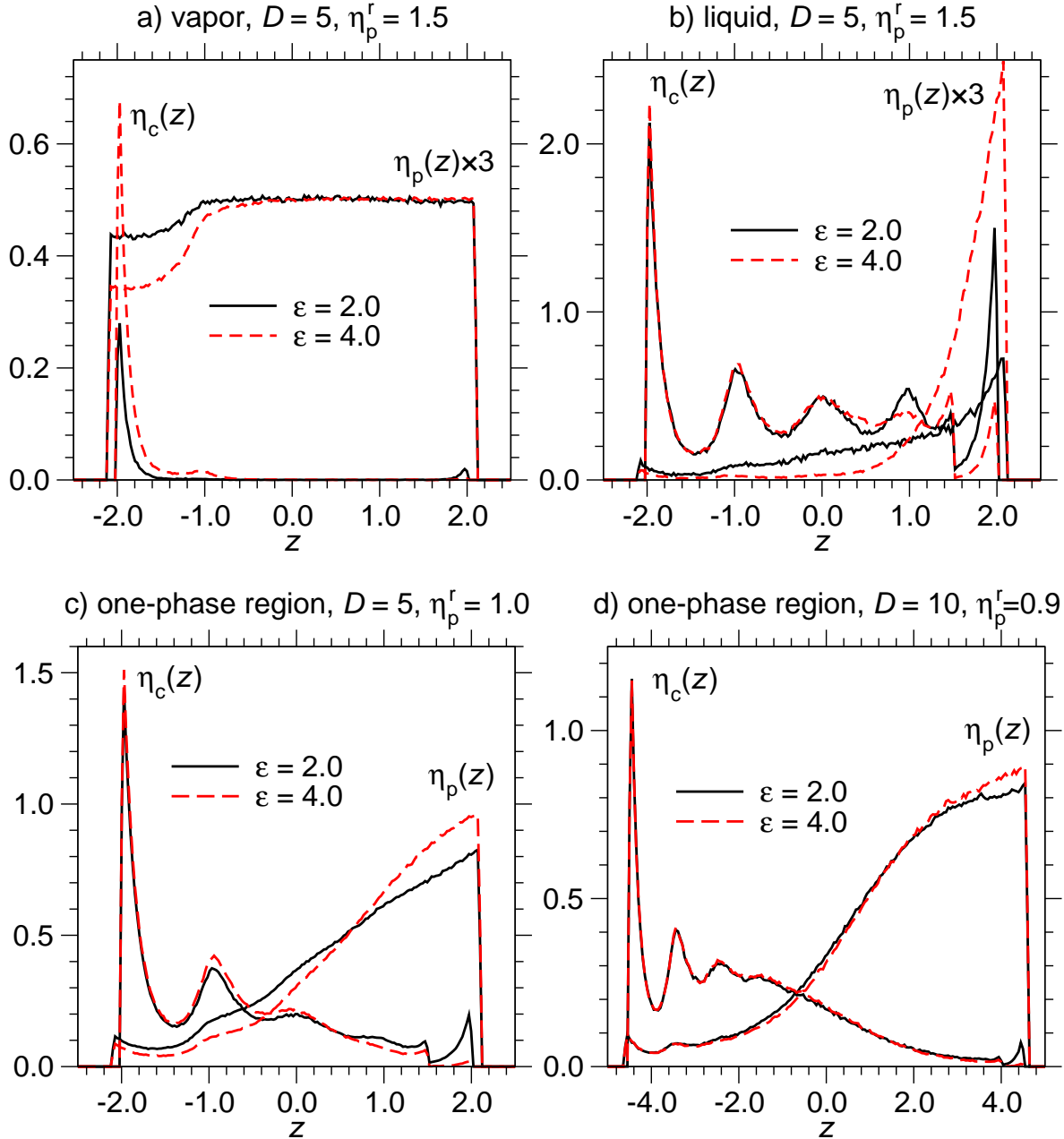

\centering
\includegraphics*[width=0.95\textwidth]{fig15a}
\vspace*{0.5cm}

\includegraphics*[width=0.95\textwidth]{fig15b}
\caption{\label{fig15} Comparison of the density profiles of colloids and
polymers in the coexisting vapor-like phase (a) and liquid-like phase
(b) between the cases $\varepsilon = 2.0$ and 4.0, for $\eta_{\rm p}^{\rm
r}=1.5$ and $D=5$. Panels c) and d) show the comparison of the density
profiles of colloids and polymers between the cases $\epsilon = 2.0$ and
4.0, but for the one-phase region in the ``soft mode'' phase. Panel c)
displays the profiles for $D=5$ and $\eta_{\rm p}^{\rm r}=1.0$, panel d)
the ones for $D=10$ and $\eta_{\rm p}^{\rm r}=0.9$.}
\end{figure}

At this point, we return to the density profiles at phase
coexistence, and compare them for the same choice of $\eta_{\rm p}^{\rm r}$
and $D=10$, but different values of $\varepsilon$, $\varepsilon = 2.0$
and $\varepsilon = 4.0$ (Fig.~\ref{fig15}). For $\eta_{\rm p}^{\rm r}=1.5$, the
vapor-like phase reaches the same polymer density for both choices
of $\varepsilon$; the main difference concerns the colloid-rich side
of the systems, the colloid enrichment at the hard wall is more
pronounced for $\varepsilon = 4.0$ than for $\varepsilon = 2.0$.
However, in the liquid-like colloid-rich phase the behavior is
just the other way round: the layered profiles of the colloid-rich
phase near the hard all are virtually identical, while the polymer
enrichment near the right wall is more pronounced for $\varepsilon =
4.0$ than for $\varepsilon = 2.0$. When one studies the effect of
varying $\varepsilon$ in the one phase region for 
$\eta_{\rm p}^{\rm r} < \eta_{\rm p, crit}^{\rm r}(D)$ however, one 
sees only a minor effect of $\varepsilon$ on the segregated structure 
where an interface has formed parallel to the walls 
(Fig.~\ref{fig15}b and \ref{fig15}d), in particular
for not extremely thin films.

\section{CONCLUSIONS}
In this paper, the Asakura-Oosawa (AO) model for colloid-polymer
mixtures for a size ratio of polymers to colloids $q=0.8$ was
studied by Monte Carlo simulation, considering thin films of
thickness $D=3$ to $D=10$ colloid diameters and
confinement between asymmetric walls. One wall is simply a
repulsive hard wall, to which the colloidal particles are
attracted via depletion forces; the other wall exerts a
square-well-type repulsive interaction (of the range of the
colloid ratio, and variable strength $\varepsilon = 0.5$ to 4.0, in
units of $k_BT=1.0$). This study complements our earlier work on
the AO model in the bulk, and under confinement between two
equivalent hard walls, where capillary-condensation like phenomena
occur; for the present model, we can smoothly interpolate from
capillary condensation-like behavior for small 
$\varepsilon$ (e.g.~$\varepsilon=0.5$ or 1.0), when both walls show some (though
unequal) surface enrichment of colloids, to interface
localization-type transitions, occurring for large $\varepsilon$
(e.g.~for $\varepsilon$ varying from $\varepsilon = 2.5$ to 
$\varepsilon=4.0$). In the latter case,
only the hard wall attracts colloids while the other wall attracts
polymers. In this region, for large $D$ the precise value of
$\varepsilon$ has little effect on the observed density profiles.
When one then increases the polymer reservoir packing fraction
$\eta_{\rm p}^{\rm r}$ (which plays an analogous role as the inverse
temperature does for thermally driven phase separation in small
molecules mixtures), one observes that the enrichment layers of
colloids and polymers at the walls gradually transform into two
domains of coexisting colloid-rich and polymer-rich phases,
separated by an interface parallel to the confining walls. We find
that the temperature dependence of the width of this interface is
considerably weaker than that of the bulk correlation length (or
``intrinsic'' interfacial width, respectively), and account for
this finding in terms of capillary wave broadening of the
interface. However, since for $D<10$ the interface profiles are
strongly affected by layering of colloids near the hard wall,
study of this broadening is difficult.

Only far away from the bulk critical point can a sharp phase
transition be observed, which we analyze by finite size scaling
methods. While for $D=3$ and not too large $\varepsilon$ the critical
value $\eta_{\rm p, crit}^{\rm r}(D)$ can be rather accurately
determined, and evidence can be found that the critical behavior
falls in the universality class of the two-dimensional Ising model, for
larger $D$ and/or larger $\varepsilon$ the Monte Carlo data are
strongly affected by problems of crossover between different
universality classes and, thus, $\eta_{\rm p, crit}^{\rm r}(D)$ can be
only estimated with rather modest accuracy, allowing no firm statements
about critical exponents. Approaching the transition
from $\eta_{\rm p}^{\rm r}<\eta_{\rm p, crit}^{\rm r}(D)$, we find a strong
increase of the correlation length $\xi_{||}$ describing the
correlation of interfacial fluctuations, but again the accuracy of
our results would not suffice to estimate the value of the
associated critical exponent. In view of the fact that even for
the simple Ising model confined between competing boundaries a
clarification of the critical behavior turned out to be very
difficult, the problems encountered for the present more
complicated model, which is strongly asymmetric even in the bulk,
are not at all surprising.

The fact that observation of interface localization does not
require very special conditions at the walls, but occurs for a
broad parameter range, is encouraging for possible experimental
tests of our results. We suggest that a repulsive interaction
acting only on the colloids could be realized by creating a wall with
a polymer brush at low grafting density.

A very interesting problem, not accessible to the present
grand-canonical Monte Carlo study, would be the dynamics of phase
separation in such a confined thin film. We hope to report on such
studies of a related model in the future.

\underline{Acknowledgments}: This work was supported in part by
the Deutsche Forschungsgemeinschaft, SFB TR6/A5 and C3.

\newpage


\end{document}